\documentclass[bibauthoryear,onecolumn]{aa}

\usepackage{hyperref}
\usepackage{mathptmx}   
\usepackage{url}
\usepackage{rotating}
\usepackage{graphicx}

\usepackage{multirow}
\usepackage{amsmath}
\usepackage{amssymb}
\RequirePackage{color}

\newcommand{\HI}{$\mathrm{H_I}$ } 
\newcommand{\Msol}{$\mathrm{M_\odot}$ } 
\newcommand{\Abell}[1]{Abell#1}
\newcommand{\NGC}[1]{NGC#1}
\newcommand{\CalibSrc}{3C161}

\newcommand{\tableline}{\hline}

\begin{document}

  \title{On sky characterization of the BAORadio wide band digital backend}
 \subtitle{Search for \HI emission in \Abell{85}, \Abell{1205} and \Abell{2440} galaxy clusters} 

\author{R. Ansari\inst{\ref{inst1}} \and
	J.E Campagne\inst{\ref{inst1}} \and
	P. Colom\inst{\ref{inst2}} \and
	C. Ferrari\inst{\ref{inst3}} \and
	Ch. Magneville\inst{\ref{inst4}} \and
	J.M. Martin\inst{\ref{inst5}} \and 	
	M. Moniez\inst{\ref{inst1}} \and
	A.S. Torrent\'o\inst{\ref{inst1},\ref{inst6}} 
}

\institute{
   Universit\'e Paris-Sud, LAL, UMR 8607, F-91898 Orsay Cedex, France
   		\& CNRS/IN2P3,  F-91405 Orsay, France\\
		\email{campagne@lal.in2p3.fr}\label{inst1}
\and
   LESIA, UMR 8109, Observatoire de Paris, 5 place Jules Janssen, 92195 Meudon Cedex, France\label{inst2}
\and
  Laboratoire Lagrange, UMR7293, Universit\'e de Nice Sophia Antipolis, CNRS, 
   Observatoire de la C\^{o}te d'Azur, 06300, Nice, France\label{inst3}
\and
  CEA, DSM/IRFU, Centre d'Etudes de Saclay, F-91191 Gif-sur-Yvette, France\label{inst4}
\and
  GEPI, UMR 8111, Observatoire de Paris, 61 Ave de l'Observatoire, 75014 Paris, France\label{inst5}
\and  
Present address: Universit\'e Paris-Sud, IPNO, UMR8608, F-91406 Orsay, France \label{inst6}
}

\today

\abstract{
We have observed regions of three galaxy clusters  at $z\sim [0.06\div0.09]$ (\Abell{85}, \Abell{1205}, \Abell{2440}), as well as calibration sources with the Nan\c{c}ay radiotelescope (NRT) to search for 21 cm emission and fully characterize the 
FPGA based BAORadio digital backend.
The total observation time of few hours per source have been distributed 
over few months, from March 2011 to January 2012, due to scheduling constraints of the NRT, which is a transit telescope. 
Data have been acquired in parallel with the NRT standard correlator (ACRT) back-end, 
as well as with the  BAORadio data acquisition system. The latter enables wide band 
instantaneous observation of the $[1250, 1500]~\mathrm{MHz}$ 
frequency range, as well as the use of powerful RFI mitigation methods 
thanks to its fine time sampling. 
A number of questions related to instrument stability, data processing and calibration are discussed. 
We have obtained the radiometer curves over the integration time range [0.01,10\ 000] seconds and we show 
that sensitivities of few mJy over most of the wide frequency band can be reached with the NRT.
 It is clearly shown that in blind line search, which is the context of \HI intensity mapping for 
Baryon Acoustic Oscillations, the new acquisition system and processing pipeline 
outperforms the standard one. 
We report a  positive detection of 21~cm emission at $3 \sigma$-level from galaxies 
in the outer region of  \Abell{85} at $\simeq 1352 \, \mathrm{MHz} \; (14 400 \, \mathrm{km/s})$ 
corresponding to a line strength of $\simeq 0.8 \, \mathrm{Jy~km/s}$. We observe also an excess power
around $\simeq 1318 \, \mathrm{Mhz} \; (21600 \, \mathrm{km/s} )$,  although at lower statistical significance,
compatible with emission from \Abell{1205} galaxies. Detected radio line emissions have been 
cross matched with optical catalogs and we have derived hydrogen mass estimates.
}

\maketitle

\section{Introduction}
\label{Intro}

%
A complete analog and digital electronic system (BAORadio), for acquisition and processing of radio signals was designed, built  by Irfu and LAL, in 2007-2009. 
The system commissioning, tests and qualification was carried out in 2009-2010 at the Nan\c{c}ay radio observatory\footnote{The Nan\c{c}ay Radioastronomy Station is part of the Observatoire de Paris and is operated by the Minist\`ere de l'\'Education Nationale and Institut des Sciences de l'Univers of the Centre National de la Recherche Scientifique.}
in collaboration with Observatoire de Paris staff. 
This development, intended for large bandwidth radio interferometers in GHz domain, 
has been achieved within the BAORadio\footnote{http://groups.lal.in2p3.fr/bao21cm/} project, 
in the context of 21 cm intensity mapping for BAO (Baryon Acoustic Oscillations) detection 
and Dark Energy science  
\citep{2012A&A...540A.129A, 2009astro2010S.234P, 2008arXiv0807.3614A, 2012IJMPS..12..256C}. 
The system has also been deployed at the CRT (Cylindrical Radio Telescope) prototype 
at CMU (Pittsburgh) in interferometric mode with up to 32 antennae \citep{2011PhDT.......158B}. Its is also being used in the PAON-4 interferometer, which is a wide band transit type radio-interferometer featuring four 5~m diameter
antennae. PAON-4 has been deployed at Nan\c{c}ay at the end of 2014 and is designed as a test bed for 3D intensity mapping.

To quantitatively qualify the performance and capabilities of BAORadio in terms of RFI filtering, sensitivities, and 
wide band blind \HI searches, the authors have used the Nan\c{c}ay Radio Telescope (NRT),
equipped with the BAORadio digital back-end and a dedicated data processing pipeline 
to search for 21 cm emission in three nearby galaxy clusters with  $0.05\lesssim z\lesssim 0.10$ 
(\Abell{85}, \Abell{1205}, \Abell{2440}), during a 11-month period, from March 2011 to January 2012.

Although our primary goal was the full and long term characterization of the BAORadio system, the target 
selection  and analysis was guided by the question of galaxy formation and evolution, and the effect of the 
environment on this processes. Obviously, knowing the atomic hydrogen mass distribution among galaxies 
and its evolution with redshift is a key element in understanding this question, as the gas reservoir 
fuels the star formation.  \HI line measurements are thus crucial in complementing multi-wavelength 
and spectroscopic data, in X-ray, UV, optical, IR , as they provide atomic hydrogen gas mass, but also 
dynamical constraints through the line width. 
Major efforts have been and are devoted to radio surveys to observe \HI mass distribution properties 
in the local universe, and its evolution with the redshift or the environment, such as the HIPASS 
\citep{2004MNRAS.350.1195M}, the ALFALFA \citep{2011AJ....142..170H}  surveys or the 
NIBLES survey at Nan\c{c}ay \citep{2009pra..confE...8V}.
%

The Arecibo Galaxy Environment Survey (AGES) \citep{2006MNRAS.371.1617A}, 
which uses the ALFA (Arecibo L-band Feed Array) has been specifically designed to 
probe \HI distribution dependence in different environment, in particular in the Virgo cluster
\citep{2012MNRAS.423..787T, 2013MNRAS.428..459T}. 

After the description of the instrument setup, observation modes and the list of observed 
targets in the next section, the data processing pipeline, RFI cleaning and calibration procedures 
are presented in Section 3. 
Section 4 is devoted to the discussion of the results obtained in term of the reached sensitivities, 
as well as some comparison with previous extragalactic \HI observations using the NRT
\citep{2003A&A...408..465M, 2001A&A...368...64V, 2002A&A...389..405P}. 
The obtained spectra and identified line emissions toward the different targets are then presented.
The cross identification with optical catalogs, using the NED\footnote{\tt http://ned.ipac.caltech.edu }
and SDSS\footnote{\tt http://www.sdss3.org/index.php} databases,  as well as measured 21 cm line 
emission parameters are discussed in Section 5. 

\section{Instrument setup and observations}
\label{Sec-the-instrument-observations}
%
\subsection{NRT optical system, receiver and correlator}
The NRT is a transit instrument of the Kraus/Ohio State design, and
consists of two mirrors, one movable and one fixed.
The tiltable primary mirror  of 40~m x 200~m (east-west) is
made up of ten flat panels (each 40m high and 20~m long) ;  the fixed spherical secondary mirror
is 300~m long and 35~m high and is located at 460~m to the South (radius: 560~m) of the primary.  
280~m North of the spherical mirror, there is a focal chariot moving
along a curved railroad track which contains a compact dual-reflector Gregorian
feed system \citep{Szymczak:2004wv, 1999JEEEA..19..111G, 1997hsra.book..229V}, with two wide band conical corrugated horns, equipped with
orthomode transducers, orthogonal linear polarization antennas and
 low-noise amplifiers, cooled to $\approx 20~\mathrm{K}$.

All celestial objects with declination greater than  $-39^{\circ}$ may be observed, 
and the west-to-east motion of the focal chariot allows for about one hour observation 
for a zero-degree declination source. The tracking time increases in proportion to the 
secant of the declination ($\propto 1/\cos \delta$).

Two receivers allow a continuous coverage of the band 1.1--3.5 GHz with optimized characteristics 
for 21 cm \HI line observations.
The low- and high-frequency receivers cover the bands 1.1--1.8 GHz and 
1.7--3.5 GHz respectively and may be used for observation of the 9 cm CH and 18 cm OH lines.
Each corrugated horn can be rotated by $\pm90^{\circ}$. Up to two linear and two circular polarizations 
can be recorded simultaneously.
The intermediate frequency (IF) bandwidth is 500 MHz. The digital correlator (ACRT) 
bandwidth can be set from 195.3 kHz to 50 MHz and has 8192 frequency channels split 
into 2 to 8 banks. Auto- and cross-correlation modes are available, and up to 4 independent 
frequency bands and/or polarization may be recorded simultaneously. 
Continuum measurements can be performed through 8 dedicated channels using the 
correlator's setup (frequency, bandwidth and polarization).


The \HI observations with the NRT are characterized by a half-power beam width (HPBW) of $3.7 \, \mathrm{arcmin}$ 
in the East-West (right ascension) direction and $22 \,
\mathrm{arcmin}$ in North-South (declination), at $\sim 1420 \, \mathrm{MHz}$ (21~cm line).
The point-source efficiency is $ 1.4~\mathrm{K/Jy}$ at zero declination $\delta = 0^\circ$ and a system temperature 
of about $T_{sys} \simeq 35~\mathrm{K}$\footnote{{\tt http://www.nrt.obspm.fr/nrt/obs/NRT\_tech\_info.html}, {\tt http://en.wikipedia.org/wiki/Nancay\_radio\_telescope}}.

We have used the following two observation modes for the program discussed here: 
\par 
- The right ascension ''Drift Scan'' mode for which the focal chariot remains stationary during each cycle, and
radio sources pass through the beam, during 180~s acquisition periods or phases. 
\par 
- ``Total power (position-switching) mode'' using consecutive cycles, each consisting of pairs 
of 40~s ON- and 40~s OFF-source integration phases, with same tracking range 
(same ground noise contribution in ON- and OFF-source spectra). \\
At the beginning of each acquisition phase, two 3-sec pulses from a noise diode (DAB) are injected at horn level to perform a relative calibration. 

\subsection{BAORadio digital back-end }
\label{BAORadio}
In this article we focus on some important features of the BAORadio system, a more complete technical description 
and its performance can be found in \citep{2011ITNS...58.1833C,2012CRPhy..13...46A}.

One important difference between BAORadio and ACRT systems is that the first one is 
embedded inside the NRT chariot as close as possible to the analog receiver, whereas in 
the second the analog signals are transferred through about 150m of cables from the chariot to the control building.

\noindent
\parbox{0.5\textwidth}{ 
The BAORadio system has been developed at LAL (Univ. Paris-Sud, CNRS/IN2P3)  \& Irfu (CEA) 
during 2007-2009 for \HI intensity mapping projects exploiting cylindrical reflectors or dense array of dishes.   
The sampling board (ADC-board) shown on the right hande side is the central component of the system. 
It can sample up 4 analog inputs at 500 Msample/s with 8 bits dynamic range and can optionally convert the 
wave forms  into frequency components through an FFT implemented on FPGA. Different firmwares allow two observation
modes: the RAW firmware can be used for waveform digitization with a maximum of 48 kSample ($96~\mathrm{\mu s}$) 
per digitization frame, while the FFT firmware performs waveform sampling, Fourier 
transform (FFT) and transmission of the Fourier coefficients (2 bytes, real/imaginary) to the acquisition computers
for 8 kSample ($16~\mathrm{\mu s}$) digitization frames, corresponding to 4096 frequency components, 
with 61~kHz resolution. 
Other digital systems with similar features such as the CASPER/ROACH 
or  UNIBOARD 
have also been developed \citep{2009astro2010T..21P}.  
}
\hfill 
\parbox{0.45\textwidth}{ 
\includegraphics[width=0.45\textwidth]{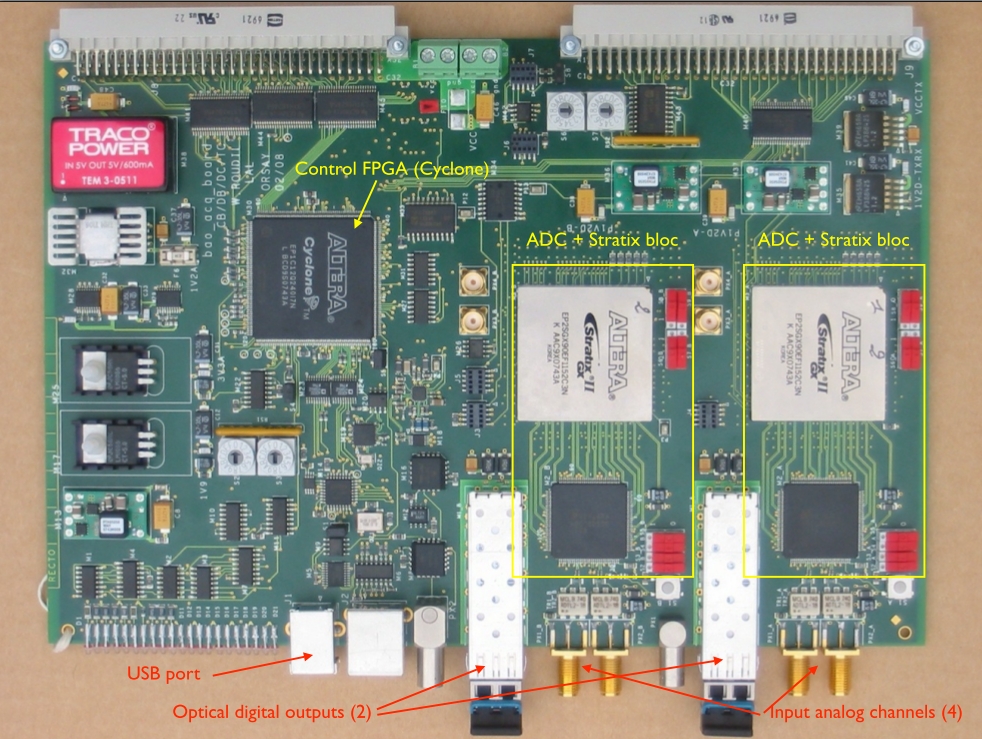}
} \\[1mm]

All observations discussed here have been performed with the RAW firmware. To cope with the acquisition system limitations, 
in term of data rate, the trigger rate was set to 8~kHz  providing 8000 digitization frames per second and leading an useful fraction of time ON-sky of 25\%. 
Each digitization frame has 16384 waveform samples, corresponding  to $\simeq 32~\mathrm{\mu s}$ signal duration. 
The 8~kHz trigger produced a total data flow dumped to disk of about 300~MByte/s (about one TByte/hour) for the two 
polarizations signals. After offline Fourier transform (FFT), we obtained 8192 complex coefficients (nicknamed a BRPaquet) corresponding 
to 30.5~kHz frequency resolution ($\simeq 6.5 \, \mathrm{km/s}$) on the entire frequency band $[1250,1500]$~MHz 
for the two polarizations.
The complete data set from all the observations have been archived at the IN2P3 computing center
using the  Irods system \footnote{ \tt http://cc.in2p3.fr/IRODS}.

\subsection{Observed targets}
Some \HI emission observation has been reported \citep{2009A&A...495..379B} for \Abell{85}, but the search of such signal has been placed in the more general context of "blind search" as for BAO intensity mapping. In this context a comparison between the 
BAORadio system and the NRT standard auto-correlator (ACRT) is reported. 

We selected a sample of 5 clusters in the redshift range $z \simeq 0.05 - 0.10$, compatible with NRT 
scheduling constraints and the BAORadio useful frequency band $\left[ 1275 , 1475 \right] \, \mathrm{MHz}$, 
corresponding to \HI velocity range $\left[-11000 , + 30 000 \right]$~km/s  for this set of pilot observations.
A first \HI study performed on two $z=0.2$ clusters with the WSRT has shown possible indications
that star-forming galaxies in the center of relaxed clusters have significantly smaller \HI masses compared
to  galaxies  located  in  the  external  regions  of  clusters,  and galaxies  located  in  merging  clusters
\citep{2007ApJ...668L...9V}. This result needs to be tested and confirmed using a larger cluster sample. 
We have thus selected merging (\Abell{85}, Abell168) and non-merging galaxy clusters (\Abell{1205}, Abell2244, \Abell{2440}) 
with a high-fraction of blue galaxies ($>20\%$ \citep{Rostagni2014}). In order to compare the gas content of star-forming
galaxies in different clusters as well as in different cluster regions, two pointings per cluster
(one within and one outside 1~Mpc from the cluster center), were originally foreseen. However, the observations 
were only carried toward three targets. 
The targets chosen are outer regions of \Abell{1205} and \Abell{85}, and the center region of \Abell{2440}. The two first have been observed much longer that the last one, so we will not report on \HI search in \Abell{2440}. The calibration sources \CalibSrc \, and \NGC{4383}\, were also included in the observation program.
The observation modes have been "total power (position-switching) mode" for the Abell clusters and \NGC{4383}, and "drift scan" for \CalibSrc.

Details on the coordinates and observation times of the targets can be found in Table~\ref{Tab:Abell}.
\begin{table*}
\small
\caption{Observation data for \HI cluster sources (Abell) and the calibration sources \CalibSrc and \NGC{4383}. 
Celestial coordinates of NRT pointing: RA and Dec, referred to J2000 epoch; 
Central observation frequency in ACRT band and observation period. Table \ref{Tab:UseFulData} gives details on analysis statistics}
\label{Tab:Abell}
\centering
  \begin{tabular}{c c c c c }
   \tableline
            Source      &  RA & Dec & $\nu_{obs}$ (MHz) & Obs. period \\ 
            \tableline
            \Abell{85}   & $00^h43^m16.99^s$ & $-9^{\circ}09^{'}46.99^{''}$ & 1346.3 & Apr. 2011 - Oct. 2011 \\ 
            \Abell{1205} & $11^h15^m08.37^s$ & $+2^{\circ}33^{'}01.39^{''}$ & 1320.8 & Mar. 2011 - Jan. 2012 \\ 
            \Abell{2440} & $22^h24^m33.30^s$ & $+0^{\circ}53^{'}18.59^{''}$ & 1302.4 & Mar. 2011 - Jun. 2011 \\ 
            \CalibSrc      & $06^h24^m43.09^s$ & $-5^{\circ}51^{'}14.00^{''}$ & 1420.4 & $9^{th}$ Dec. 2011   \\     
            \NGC{4383}    & $12^h25^m25.5^s$  & $+16^{\circ}28^{'}12^{''} $  & 1412.5 & $22^{th}$ Oct. 2011 \\      
  \tableline
   \end{tabular}
\end{table*}
\Abell{85} is a rich cluster located at $z \simeq 0.055$ and have been extensively studied in optical \citep{1998A&AS..128...67S, 1998A&AS..129..281D} and  in X-rays \citep{1998A&A...335...41D, 2001A&A...368..440L, 2010PASJ...62..743T}.
This cluster has also been studied in 21 cm, using VLA and observation of \HI emission of some of the blue 
galaxies in the outer region has been reported  \citep{2009A&A...495..379B}. However, no detail or 
figures quantifying the emission strength of velocity/redshift from these galaxies can be found in the 
literature \citep{2008pc1..conf..102B}. 

\begin{figure}[htbp]
\centering
\begin{minipage}{0.45\textwidth}
\includegraphics[width=0.95\textwidth]{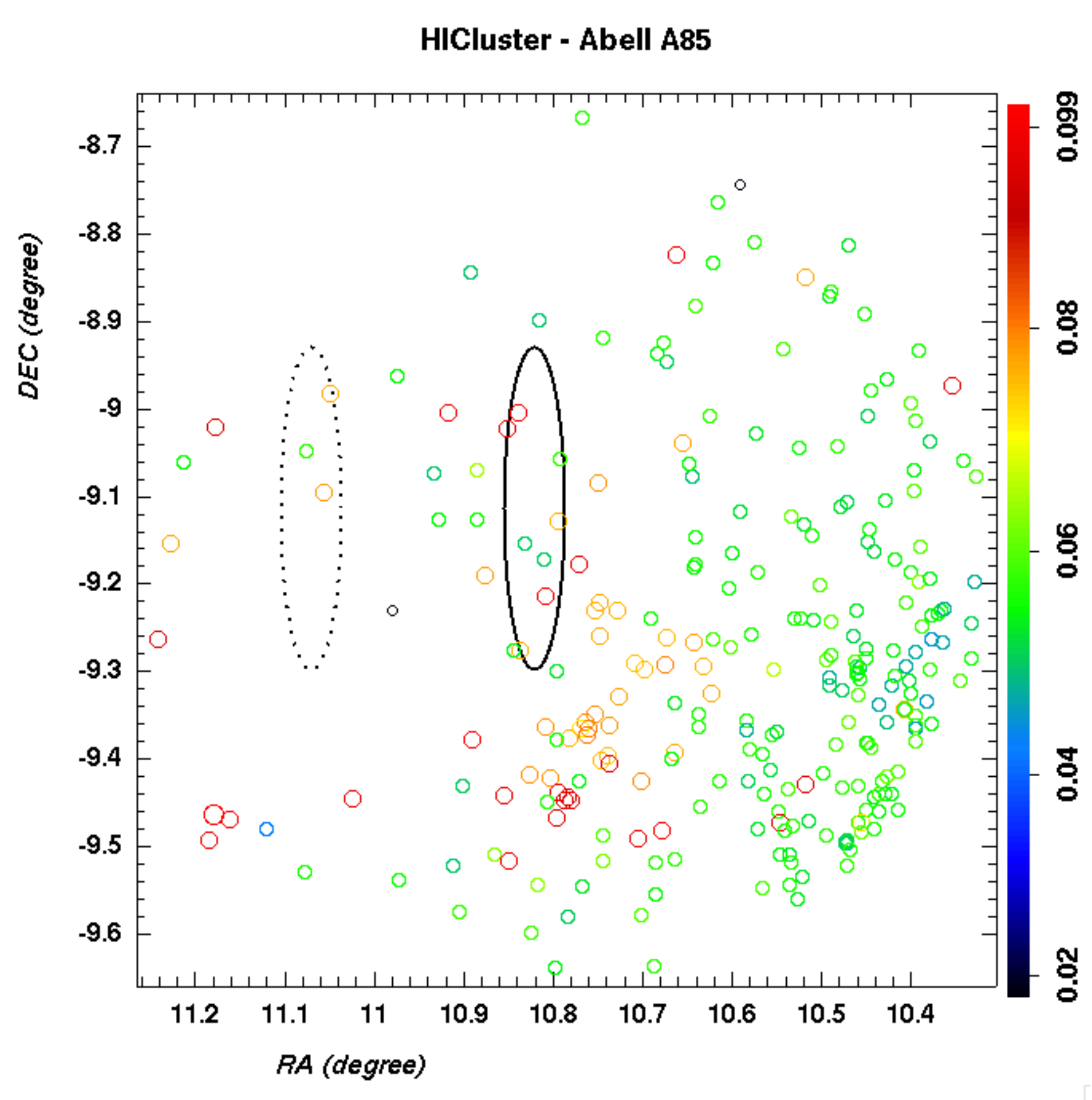}
\caption{Distribution of optical galaxies in the redshift range $0.02<z<0.1$ 
around the NRT ON/OFF beam positions near \Abell{85} cluster in the (RA, DEC) plane.
Color scale represents the galaxy redshifts, NRT ON and OFF beam spots are 
represented as solid and dashed ovals}
\label{Fig-A85-gals-z}
\end{minipage}
\hspace{2mm}
\begin{minipage}{0.45\textwidth}
\includegraphics[width=0.95\textwidth]{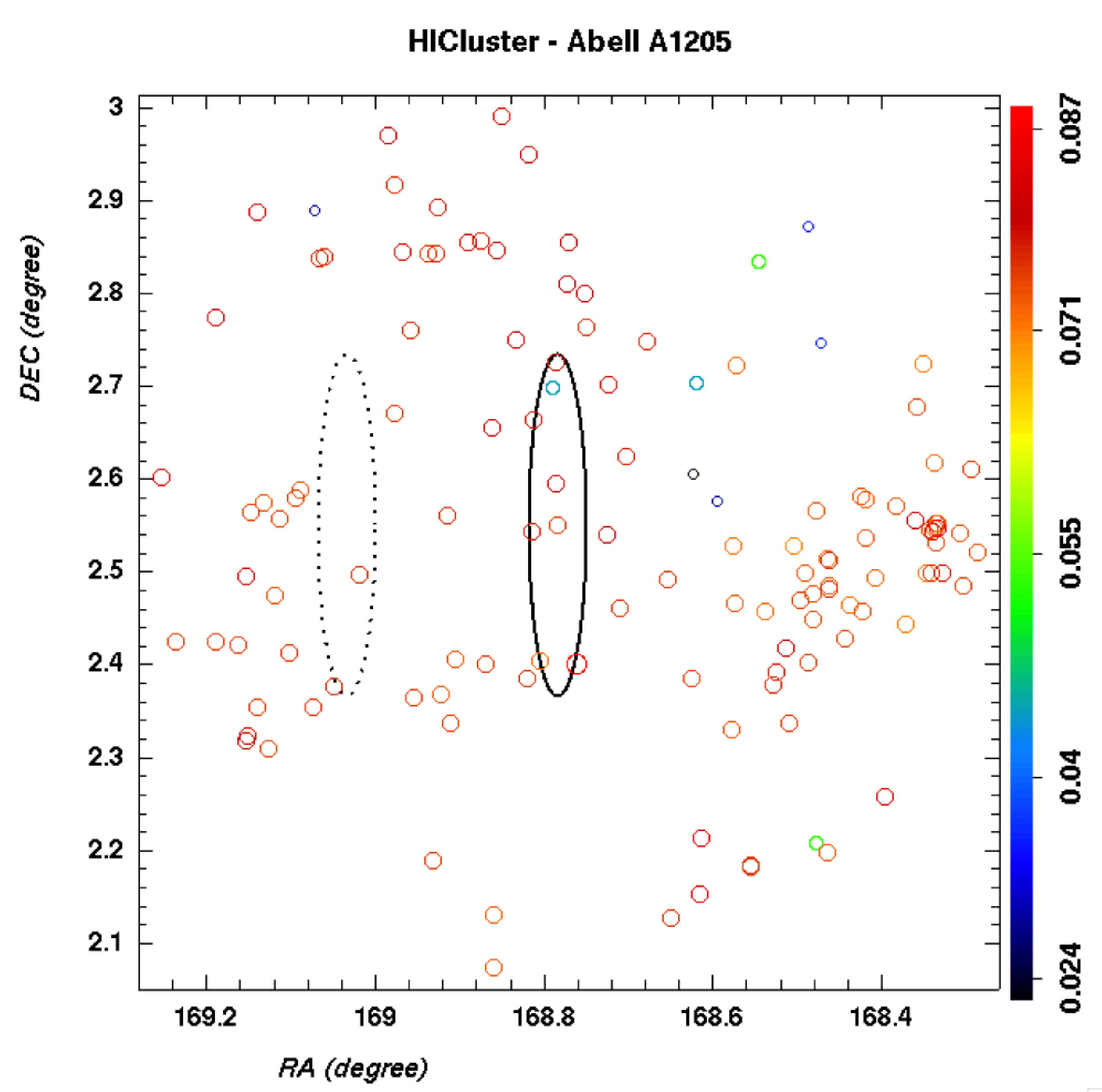}
\caption{Distribution of optical galaxies in the redshift range $0.02<z<0.1$ 
around the NRT ON/OFF beam positions near \Abell{1205} cluster in the (RA, DEC) plane.
Color scale represents the galaxy redshifts, NRT ON and OFF beam spots are 
represented as solid and dashed ovals}
\label{Fig-A1205-gals-z}
\end{minipage}
\end{figure}

The spatial distribution of optically identified galaxies with spectroscopic redshifts near the target position 
observed in this program is shown  in Fig.~\ref{Fig-A85-gals-z}. The spectroscopic redshifts are represented 
using the color scale, the ON and OFF NRT beam positions and footprints are also shown. 
The \Abell{1205} has not been as extensively studied; the optically selected galaxy distribution around the observed \Abell{1205} targets 
is shown in Fig.~\ref{Fig-A1205-gals-z}.

\section{Data analysis}
\label{Sec:DataAnalysis}
Two data processing pipelines have been setup to handle data from the BAORadio system, and 
the standard NRT correlator (ACRT) system, which have similar stages. 
An overview of the BAORadio pipeline is presented in the next section, followed by the calibration procedure
that has been used. Some additional details are given in appendix \ref{App:procpipeline}, and  
the ACRT data reduction pipeline is also briefly described in appendix \ref{App:ACRT}.

\subsection{BAORadio data reduction pipeline}
Our data reduction pipeline features the following stages:
\begin{enumerate}
\item We first perform an FFT on each digitization frame of 16384 ADC samples, 
yielding 8192 complex Fourier components, distributed 
over the $\left[ 1250,1500 \right] \mathrm{MHz}$ total bandwidth. Frequency 
components have thus a resolution of $\simeq 30.5~\mathrm{kHz}$ or $\simeq 6.4~\mathrm{km/s}$.
Each digitization frame and hence each spectrum is time-tagged with $8~\mathrm{ns}$ precision by the 
ADC-board firmware. However we consider that our absolute time (UT) precision, 
given by the acquisition computer clock, is around 1-sec.  
Using the scheduling log files of the NRT operation, these time-tags are used 
to extract data sets corresponding to different operation sequence of the NRT: diode pulses, focal chariot 
motion, plane mirror motion \dots\ . Notice that these sequences differs for the 
"Position-Switching" (PS) observation mode  used for the three clusters, or  the "Drift Scan" (DS) mode
used for \CalibSrc\, (see  Sec.~\ref{Sec-the-instrument-observations}). 
As each phase starts with a DAB pulse (see Sec.~\ref{Sec:Calibration}) and stops with NRT chariot positioning movement, some margin has been taken to allow for system stabilization, so we use only 30-sec data over the 40-sec 
available per ON and OFF phases of a cycle.
\item RFI filtering.  \\
Outside the radio protected band [1400,1427] MHz, many terrestrial signals from communication systems, 
radars\ldots create spurious signals in the surveyed frequency band [1250,1500] MHz at Nan\c{c}ay. 
We have used the very fine time sampling of the BAORadio system ($ \lesssim  0.1 \mathrm{ms}$), 
to implement an RFI mitigation method for fast intermittent signals, such as radar pulses, 
using median filtering over time, for each frequency component.  
Figure \ref{Fig-BAOelec-RFI-1} shows for instance a time-frequency map, corresponding to $\simeq 86~\mathrm{s}$ observations of \Abell{85}, with $0.6~\mathrm{s}$ time resolution and $\simeq 30.5~\mathrm{kHz}$ frequency resolution. 
A number of intermittent and permanent RFI signals are clearly visible. We have applied our RFI cleaning 
method independently to each polarization signal. 
A more complete description of our RFI mitigation method and corresponding results, 
applied to each polarization signal $p0, p1$ independently, is given in appendix (\ref{App:RFIFilter}).
 
\begin{figure}
\centering
\includegraphics[width=0.75\textwidth]{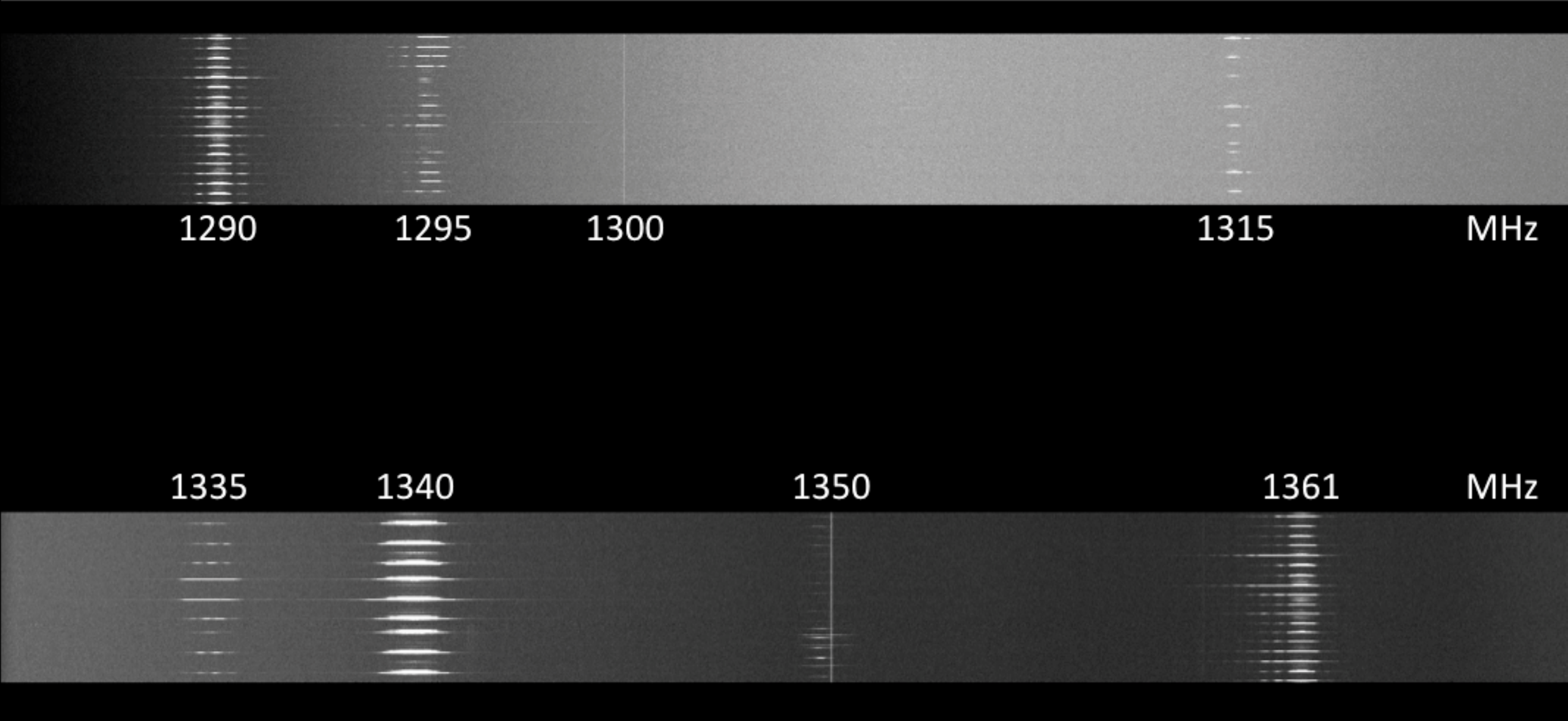}
\caption{Time-frequency map of $\simeq 86 \mathrm{s}$ observations of \Abell{85} for one polarization
showing intermittent  strong radar signals and permanent narrow RFI signals originating from electronic components. The frequency bands covered are $[1285, 1325]$~MHz (top) and $[1330, 1370]$~MHz (bottom). The other polarization (not shown) experiences the same features}
\label{Fig-BAOelec-RFI-1}
\end{figure}

\item Rejection of noisy observations. \\
These RFI cleaned spectra are then used to compute average spectra for each DAB, ON, OFF 
phases of each cycle, for each polarization signal $p0, p1$. 
These cycle/cycle spectra are then used to produce average spectra over different 
coarser time scales, per run (observations during a given day), or the whole period. 
We apply however some additional cuts to delete noisy observations, affected by the Sun 
transits for instance, or a small fraction of the data affected by electronic instabilities. 
We find that the BAORadio system associated with the NRT has been quite stable,
and reliable; we have been able to keep more than $95 \%$ of the whole data sets 
for all Abell clusters for the final analysis, compared to 50-60\% for standard 
NRT ACRT data  (Table~\ref{Tab:UseFulData}). 

\item Correction to the Local Standard of Rest (LSR) frame. \\
For each observation run (day), we have computed the doppler shift due to the movement of the Sun with respect to nearby stars, and corrected the corresponding spectra to obtain spectra in the LSR reference frame. The correction is applied separately for each cycle, 
polarization ($p=0,1$), and On/Off-source ($s=ON,OFF$) spectra described above. 
Notice that the frequency shifts are $\lesssim 140~\mathrm{kHz}$, compared to our $ 30~\mathrm{kHz}$ frequency resolution;
we do not do any interpolation along the frequency axis and spectra are only shifted according to an integer rounded number of frequency bins. 

\item Computation of the overall frequency response $g(\nu)$. \\
The spectral response $g(\nu)$, represents  the NRT optics and the BAORadio 
electronic chain frequency response, although the variations are mainly due to 
the analog electronic filter shape.
We determine a normalized frequency response $g(\nu)$ using measured spectra $P(\nu)$, 
after filtering and smoothing, so that\\  $P(\nu)/g(\nu) \simeq 1$. 
We have developed two analysis methods, starting from the same 
set of cycle/cycle spectra, including the frequency response determination $g(\nu)$, 
adapted for the two  calibration methods, using either \CalibSrc\, radio source, or the Milky Way 1420 MHz line. \\

(a) For each cycle, a frequency response function is computed from a smoothed, 
median filtered version of OFF-source spectra, for each polarization signal\\ $g_{0,1}(\nu) = \mathrm{Off}_{filt}$. 
The final signal toward a given target is obtained by averaging, over all cycles, 
the difference of ON-source and OFF-source  spectra, renormalized by the gain, for each polarization.
$$ \langle \mathrm{(On-Off)/Off_{filt}} \rangle = \left\langle \frac{ P_{0,1}^{On} - P_{0,1}^{Off} }{g_{0,1}(\nu)} \right\rangle_{cycles} $$ 

(b) Average ON-source and OFF-source spectra are computed for each run (average of all cycles for a given day), 
and are then used subsequently to compute a normalized frequency response for each run $r$, 
each polarization $(p=0,1)$, for the ON-source and OFF-source spectra separately $g_{r,(0,1)}^{On,Off}(\nu)$. 
The average per cycle, per polarization ON/OFF spectra are first filtered with a 
sliding median filter over a window of $15\, \mathrm{MHz}$ wide, and then modeled as a spline 
curve with $4\, \mathrm{MHz}$ step. The smooth spline frequency responses  $g_{r,(0,1)}^{On,Off}(\nu)$
are free from RFI and are used to correct the original ON/OFF-source spectra, for each run and polarization.
The final signal toward each observed ON-source and OFF-source is then computed by averaging 
the normalized gain corrected spectra from each observation run:
\begin{eqnarray*}
\langle \mathrm{On / g_{On} } \rangle & = & \langle  P_{0,1}^{On} (\nu)  \, / \, g_{r,(0,1)}^{On} (\nu)  \rangle_{runs}  \\
\langle \mathrm{Off / g_{Off} } \rangle & = & \langle P_{0,1}^{Off} (\nu)  \, / \, g_{r,(0,1)}^{Off} (\nu)  \rangle_{runs}  
\end{eqnarray*}
\item Radiometric calibration. \\
The final average spectra obtained in the previous step are not expressed in physical units, 
but in r.a.u (Relative Arbitrary Unit). 
The conversion factor into physical units, brightness temperature (Kelvin) or source intensity (Jansky) 
is determined using the \CalibSrc\, source for the (a)  $ \langle \mathrm{(On-Off)/Off_{filt}} \rangle $ spectra and 
the Milky Way 1420 MHz line for (b) $\langle \mathrm{On / g_{On, Off} } \rangle$. The calibration procedure 
is described in next section. 
\end{enumerate}

We obtain fully compatible final spectra, using the two methods (a), (b) described above to determine 
frequency response and compute radiometric calibration coefficients. This will be discussed further in Sec.~\ref{Sec:results}.  

\subsection{Residual modulations in  the final spectra}
\label{Sec:Oscillations}
We observe residual structures in the ON-source or OFF-source spectra, independent of the processing method used (a) or (b).
As shown  in Fig.~\ref{Fig-Oscillations3} we have observed two regimes of modulation of the spectra. The first one, with a particularly unexpected large amplitude of $\gtrsim 100~\mathrm{mJy}$, has a modulation period in frequency of about $3-4 \, \mathrm{MHz}$. It is attributed  to standing waves in the $\simeq 25~\mathrm{m}$-long cables connecting NRT front-end electronics and the BAORadio analog chain, combined with a slight impedance mismatch in the BAORadio analog electronic input stage.
The second one, smaller in amplitude with $\simeq 500 \, \mathrm{kHz}$ period structures in frequency, (see zoom in Fig.~\ref{Fig-Oscillations3}) is attributed to the standing waves between the large spherical mirror and the horn, located $\simeq 280~\mathrm{m}$ north of the spherical mirror. 
Notice that this last modulation had suffered from a sudden $\pi/2$ phase shift during a period of cycles. 
Unfortunately both kind of modulations not only are non uniform on the total 250~MHz bandwidth but also are time varying. Different kind of filtering have been tried to analyze the ON or OFF spectra alone without any success and it is the reason why we have used (ON-OFF) spectra which reduce considerably these modulations as one can appreciate looking at cyan versus blue spectra in Fig.~\ref{Fig-Oscillations3}. 

%

\subsection{Radiometric calibration}
\label{Sec:Calibration}

\begin{figure}
\centering
\begin{minipage}{0.55\textwidth}
\includegraphics[width=0.95\textwidth]{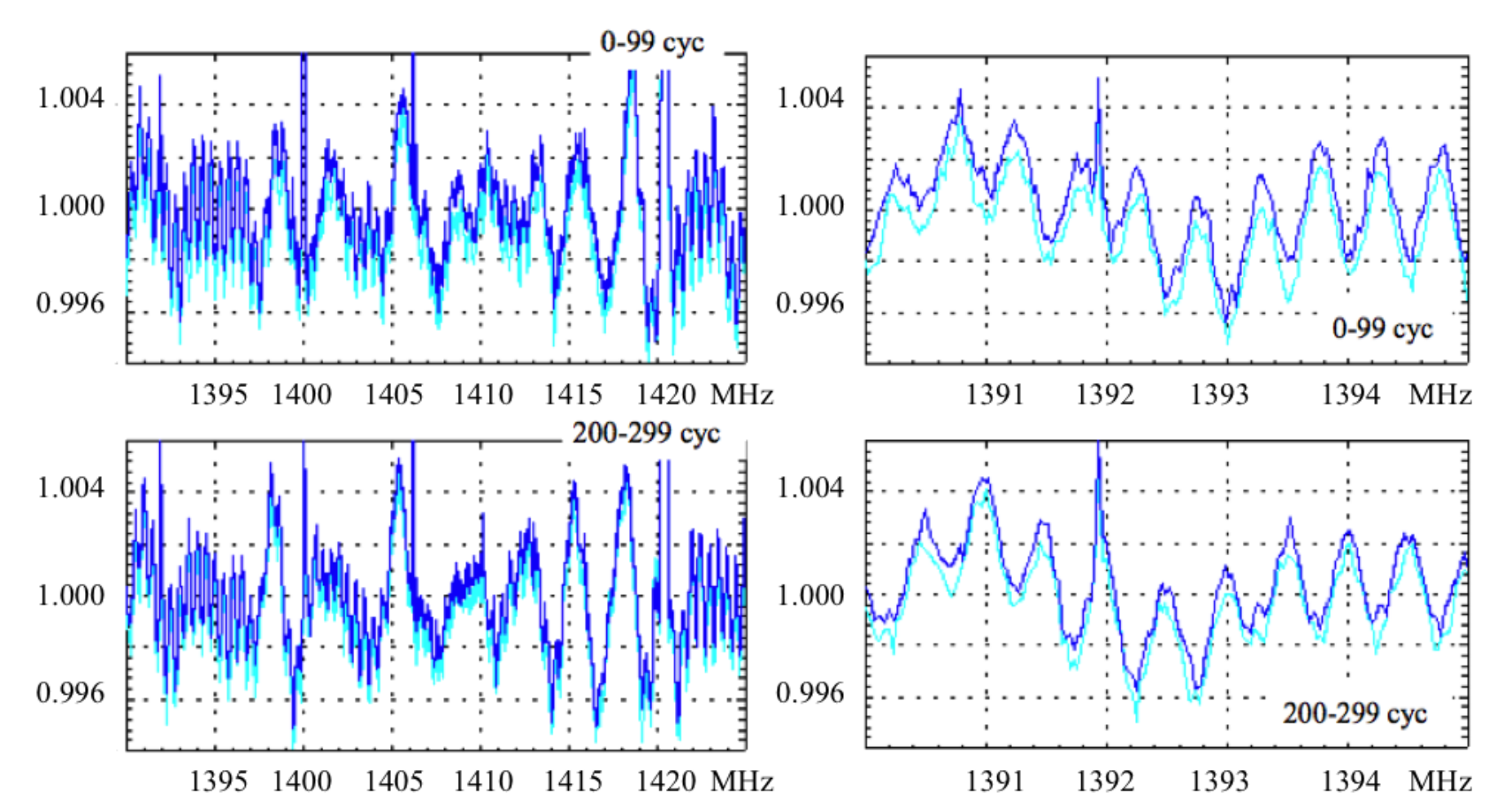}
\caption{Left panels: cumulated On/Off$_{filt}$ (cyan) and Off/Off$_{filt}$ (blue) for polarization p0, segmented into two periods of 99 cycles from the observations of \Abell{1205}. Clear modulations of the spectra are visible with a frequency of the order of 3-4~MHz. Right panels: a zoom of the left panels clearly shows 500~kHz modulation. Notice that between the two right panels the modulation phase has changed by $\pi/2$}
\label{Fig-Oscillations3}%
\end{minipage}
\hspace{4mm}
\begin{minipage}{0.4\textwidth}
\includegraphics[width=0.95\textwidth]{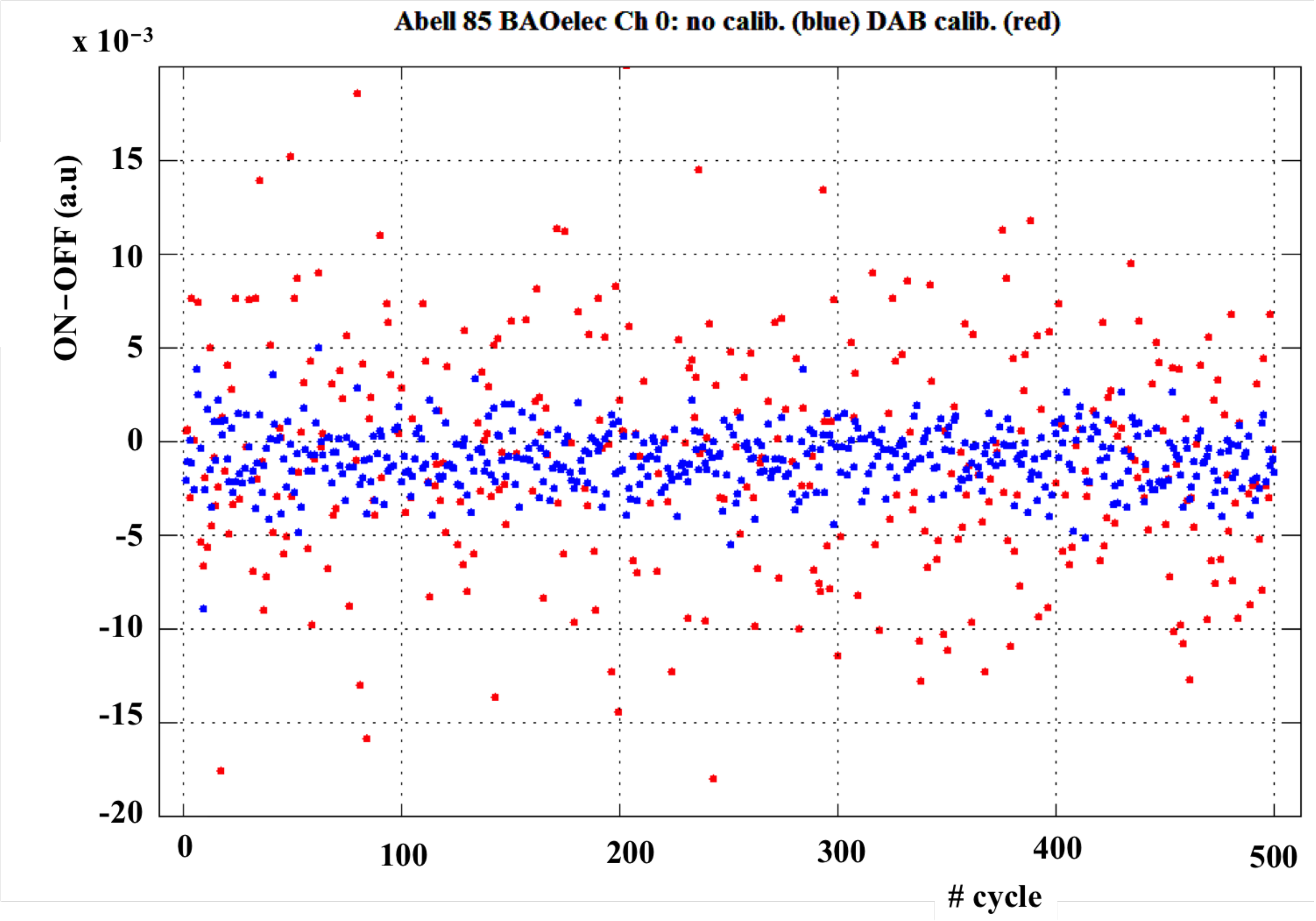}
\caption{Distributions of the difference of power (ON-OFF) integrated in the frequency band $[1420.2, 1420.6]$MHz (arbitrary unit) around the Milky Way \HI line for one polarization (polar.~0) given by the BAORadio system. The blue dots are obtained with no calibration applied while the red dots show the effect of the DAB calibration applied for each cycle (ON, OFF). The use of the DAB calibration increases the sigma of the ON-OFF by a factor approximatively $2.5$}
\label{Fig-BAOelec-DABnoDAB}
\end{minipage}
\end{figure}

The standard procedure provided by the NRT for calibration is based on the use of noise diodes (DAB) signal injected at the horn level. Initially, we followed the NRT procedure 
for relative calibration using the DAB signals, but we have shown that this method induces additional instabilities, leading to a significant increase of the final signal fluctuations or noise level. 
Figure \ref{Fig-BAOelec-DABnoDAB} shows the increase of fluctuations where DAB signal is used for relative cycle/cycle signal level calibration. The same effect is observed for the NRT ACRT signal and discussed in appendix \ref{App:ACRTCalibration}. We have found that the NRT and BAORadio chains are quite stable, even over durations of a few months, and we have only used the radiometric sources and the Milky Way \HI line to determine the absolute calibration.

\subsubsection{Radiometric calibration using \CalibSrc}
\label{Sec:Calibration_3C161}
The radio-continuum source \CalibSrc\, has been observed in drift scan mode. During the entire observation time, we have registered several \CalibSrc\, transits with both polarizations. The average of the different transits is shown in Fig.~\ref{Fig-BAOelec-3C161-DS-all}. 


%



We have derived the absolute radiometric calibration coefficient $C_a$ for the average spectra\\  
$ \langle \mathrm{(On-Off)/Off_{filt}} \rangle $ by comparing the peak value of \CalibSrc\, transit, after baseline subtraction, averaged 
over the 2 polarizations, 
to the published emission intensity measurement for \CalibSrc\, \citep{1994A&A...284..331O,1977A&A....61...99B} of 
$18.58\pm0.09$~Jy 
for the continuum flux collected by a circularly-polarized total power receiver centered at 1408~MHz, with a 20~MHz bandwidth ([1398, 1418]~MHz). 
The quoted error is a systematic error reflecting the 3\% difference between the two referenced measurements. It yields
$$ C_a = 22.6 \pm 0.6 \, \mathrm{Jy / r.a.u} $$ 
The radiometrically calibrated spectra in Jansky corresponding to the difference of ON-source, OFF-source for 
each cluster are obtained by multiplying the average of the two polarizations by $C_a$:
$$ S(\nu) = \frac{C_a}{2} \left( \langle \mathrm{(On-Off)/Off_{filt}} \rangle_{p0} + \langle \mathrm{(On-Off)/Off_{filt}} \rangle_{p1} \right) $$

Using the spectral fit of reference \cite{1994A&A...284..331O}, we have determined the $C_a$ coefficients for the frequency band covered by the three Abell clusters. The coefficient is constant at a few percent level, we are thus confident that our \CalibSrc\, calibration method is valid indeed for the whole frequency band needed to search for \HI emission line (Sec.~\ref{sec-hi-signal}).


The above value of the derived calibration coefficient has also been checked using the \HI source \NGC{4383}. See the appendix for more 
details.


\subsubsection{Radiometric calibration using Milky Way 1420~MHz line}
\begin{figure}[htbp]
\centering
\begin{minipage}{0.47\textwidth}
   \includegraphics[width=0.95\textwidth, height=0.7\textwidth]{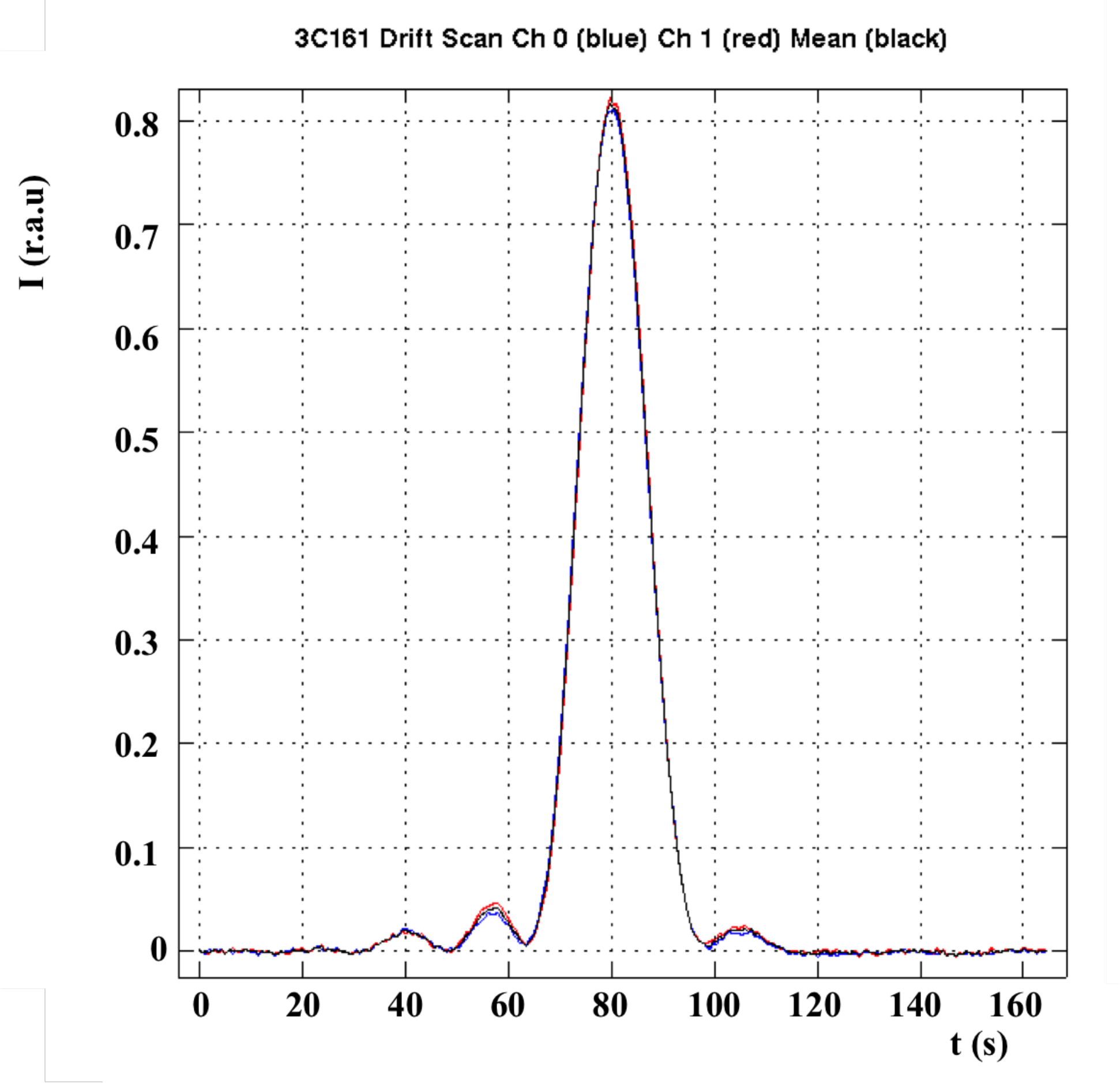}
   \caption{BAORadio mean $\langle\mathrm{(On-Off)/Off}_{filt}\rangle$ integrated in the [1412, 1413]~MHz band averaged over the transits of \CalibSrc\, observation. Both polarization channels (red/blue) as well as their mean value (black) are displayed}
   \label{Fig-BAOelec-3C161-DS-all}%
\end{minipage}
\qquad
\begin{minipage}{0.47\textwidth}
\includegraphics[width=0.95\textwidth]{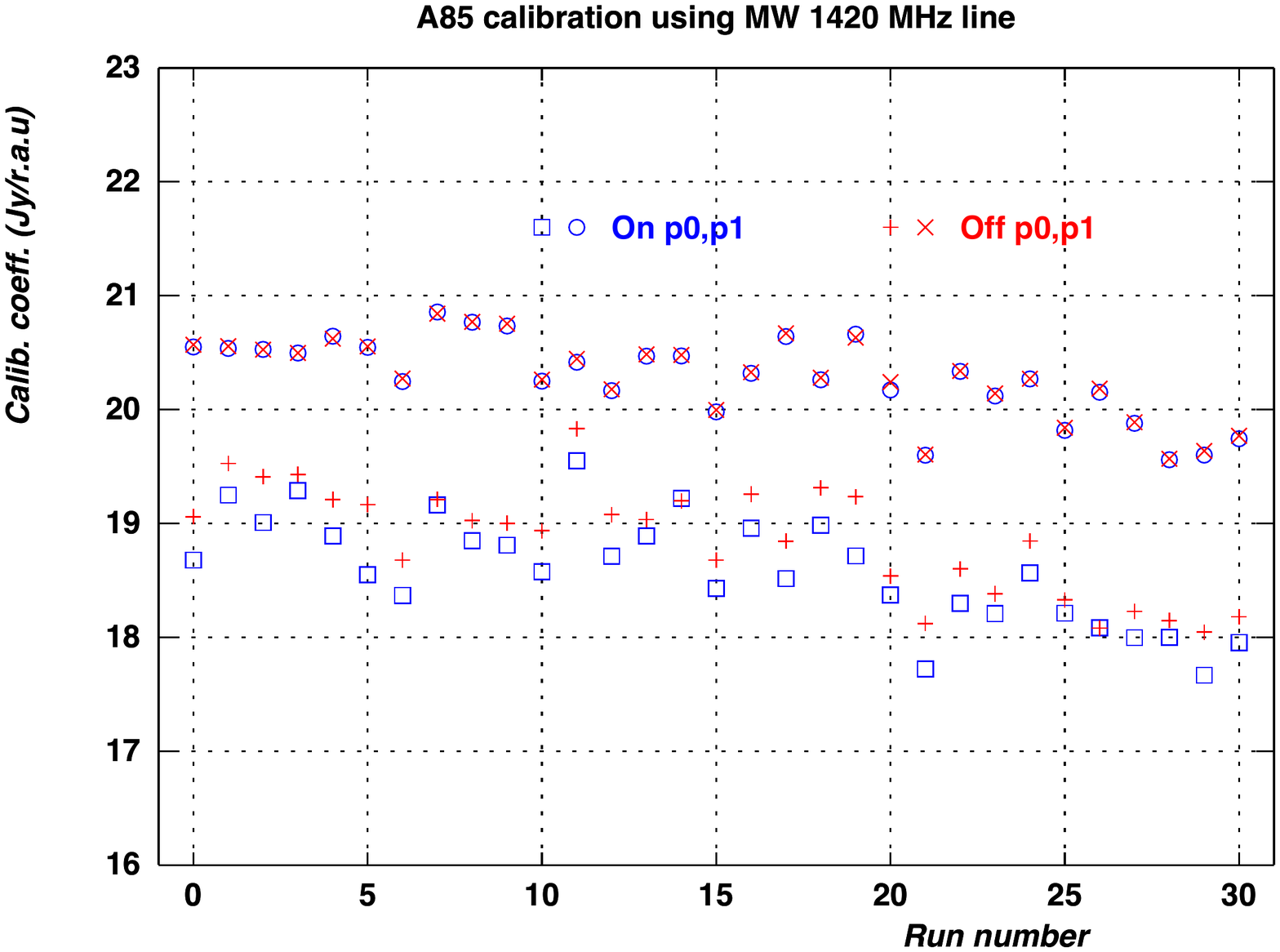}
\caption{variation of the \Abell{85} calibration coefficients for the $32$ used observation runs. 
Blue squares and circles corresponds to the two polarizations ($0,1$) and the ON-source pointings, while red plus and cross markers corresponds to the OFF-source pointings}
\label{Fig-vcalhi}
\end{minipage}
\end{figure}

The frequency band observed contains the Milky Way \HI emission line at 1420~MHz.
Using the ON-source and OFF-source observations separately we have identified the Milky Way emission line, 
corrected from Doppler shift (i.e. LSR correction), and compared its strength to the Leiden/Argentine/Bonn (LAB) Survey of Galactic \HI database \citep{2005A&A...440..775K}.
This procedure is indeed  quite difficult as the ON and OFF spectra are affected in a similar way by the residual 
structures in the spectra
(see Sec.~\ref{Sec:Oscillations}). 
To minimize the impact of these residual structure, we have performed a spline modeling of the base line, before extracting the Milky Way
emission line strength from our ON,OFF-source spectra. Notice that the $\langle\mathrm{(On-Off)/Off}_{filt}\rangle$  
spectra obtained with the  (a) method, computed cycle by cycle automatically cancel
the above mentioned artifacts.
We have also checked that the temperature measured by the LAB survey doesn't vary significantly in the NRT beam.
The LAB survey measurement are expressed as brightness temperature, so we have used the 
standard NRT antenna temperature $T_a$ to point source intensity $I_s$ conversion factor, as given by 
the NRT point source efficiency of $\eta = 0.56$ and a collection area 
$A = 200 \times 35 = 7000 \mathrm{m^2}$  (G\'erard~E., private communication). It yields 
$$ 2 \, k_B \, T_a = \eta  \, A \, I_s \hspace{3mm} \rightarrow \frac{T_a}{I_s} = \frac{\eta \, A}{2 k_B} = 1.42 \, \mathrm{K/Jy} $$ 
We have computed an absolute radiometric calibration coefficient $C_b$, for each polarization signal, 
each observation run (day) and ON-source, OFF-source spectra, and for each observed cluster.   
The calibration is rather stable: the figure \ref{Fig-vcalhi} shows the variation of the \Abell{85} calibration coefficients
for the $32$ runs over a $4$ months period from April to October 2011.
The derived value of the calibration coefficient, with a systematic error of 5-10 \%, and statistical error 
of a few percent: 
$$ C_b \simeq 20 \pm 2 \, \mathrm{Jy / r.a.u} $$ is compatible with the $C_a$ coefficient obtained from the \CalibSrc\, 
source.

%
\section{Results}
\label{Sec:results} 
\subsection{Sensitivity as a function of integration time}
\label{Sec:sensitivity}
\begin{figure}
\centering
\includegraphics[width=0.75\textwidth]{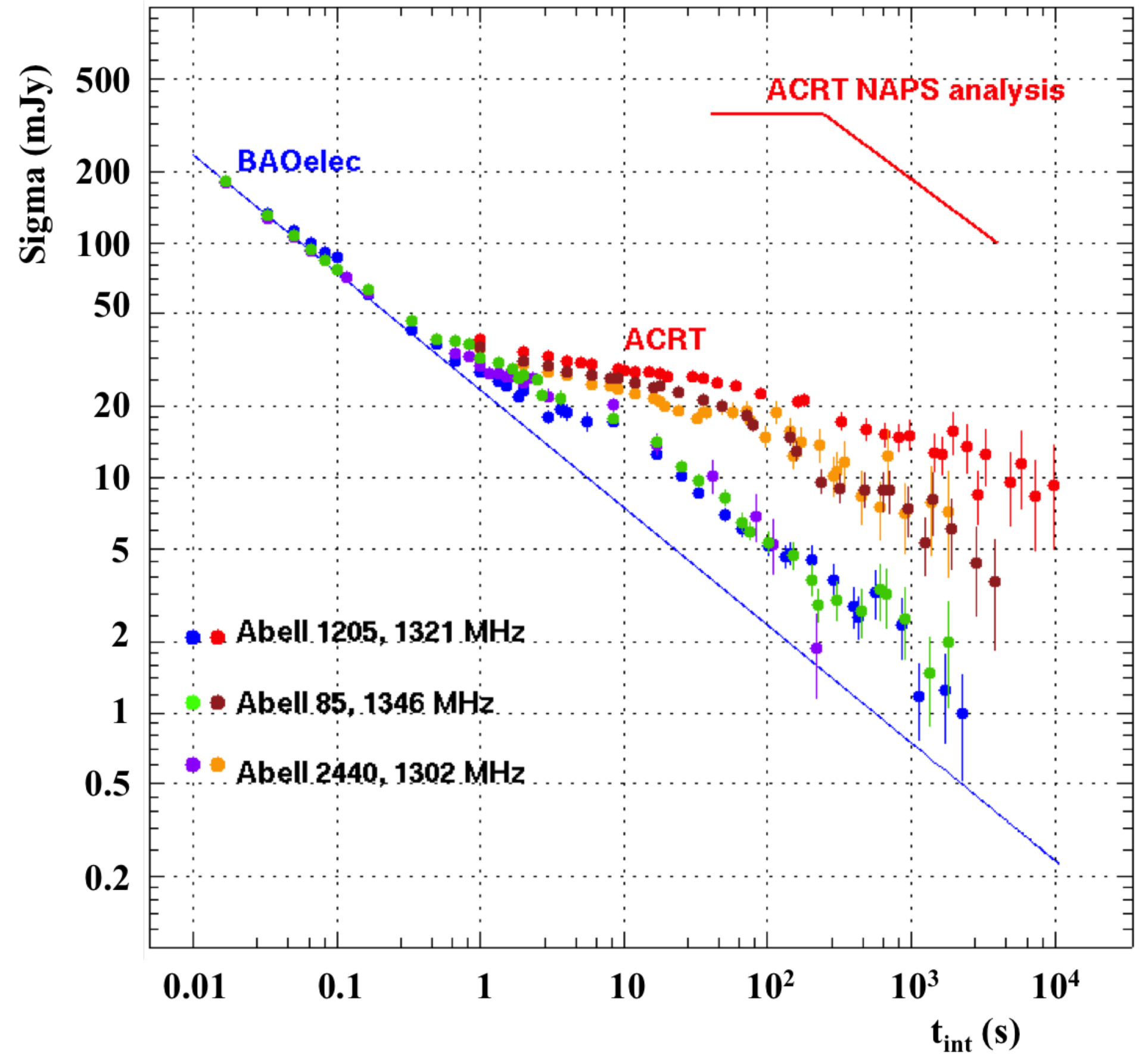}
\caption{Radiometer curve, sigma of the mean of the two polarization signals averaged over 1~MHz around the central frequency observed in ACRT versus integration time, obtained for BAORadio (ACRT) data for the three clusters: \Abell{1205} in blue (red) around 1321~MHz, \Abell{85} in green (brown) around 1346~MHz and \Abell{2440} in purple (orange) around 1302~MHz. The blue solid line represents the expected curve for a system affected only by white noise. The red line represents the curve obtained for all clusters with ACRT data analyzed with the NRT standard pipeline. The sigma of the last points are estimated with 5 samples of the quoted integration time. The total statistics on each Abell cluster are quoted in Table~\ref{Tab:UseFulData}}
\label{Fig-Abell1205-2440-85-Sensit}%
\end{figure}

For each cluster observation, we have applied the RFI filtering and data cleaning procedures exposed in previous sections, then we have used the time series of  \mbox{$\langle\mathrm{(On-Off)/Off}_{filt}\rangle$} spectra integrated over 1~MHz around the ACRT observation frequency (Table~\ref{Tab:Abell}) to draw for both ACRT and BAORadio systems the evolution of the fluctuation R.M.S
as function of the integration time, the so-called \emph{radiometer curve} (details on the ACRT data analysis are in Sec.~\ref{App:ACRT}). To compare the two systems, we have defined an "effective integration time" taking into account the difference of useful fraction of time ON-sky observation: 100\% for ACRT and 25\% for BAORadio. The result is shown in Fig.~\ref{Fig-Abell1205-2440-85-Sensit} for BAORadio (reap. ACRT): \Abell{1205} in blue (red) around 1321~MHz, \Abell{85} in green (brown) around 1346~MHz and \Abell{2440} in purple (orange) around 1302~MHz. The blue solid line represents the expected curve for a system affected only by white noise. The red line of Fig.~\ref{Fig:NAPS} (top) represents the curve obtained for all clusters with ACRT data analyzed with NRT standard NAPS pipeline \citep{NAPSUserGuide}.



The ACRT data has been processed down to its minimum integration time of 1-sec. For BAORadio, we have processed part of the data down to 16.7~ms integration time and the whole data set at 8.4~sec.
In the range of integration time greater than 1-sec, the ACRT sigma is larger by factor two than the BAORadio one. Moreover the ACRT sigma evolution differs from a white noise trend which is a sign of additional noises present in the integrated $\langle\mathrm{(On-Off)/Off}_{filt}\rangle$ in the frequency band of interest. This is due to both differences between ON and OFF power levels within the same cycle and from cycle to cycle variations. 
Notice that our ACRT pipeline gives notably better results than the standard NAPS pipeline, when we consider the total power 
evolution during cycles, as shown in Fig.~\ref{Fig:NAPS} (top).

For BAORadio the radiometer curve is similar to a pure white noise below 1-sec and remains lower than the ACRT curve for greater integration times. One possible origin of the "1-sec" trend change is the noise generated by the 1-sec duty cycle of the cryogenic cooling system for the low noise amplifier in the chariot. 



\begin{figure}
\centering
\includegraphics[width=0.55\textwidth]{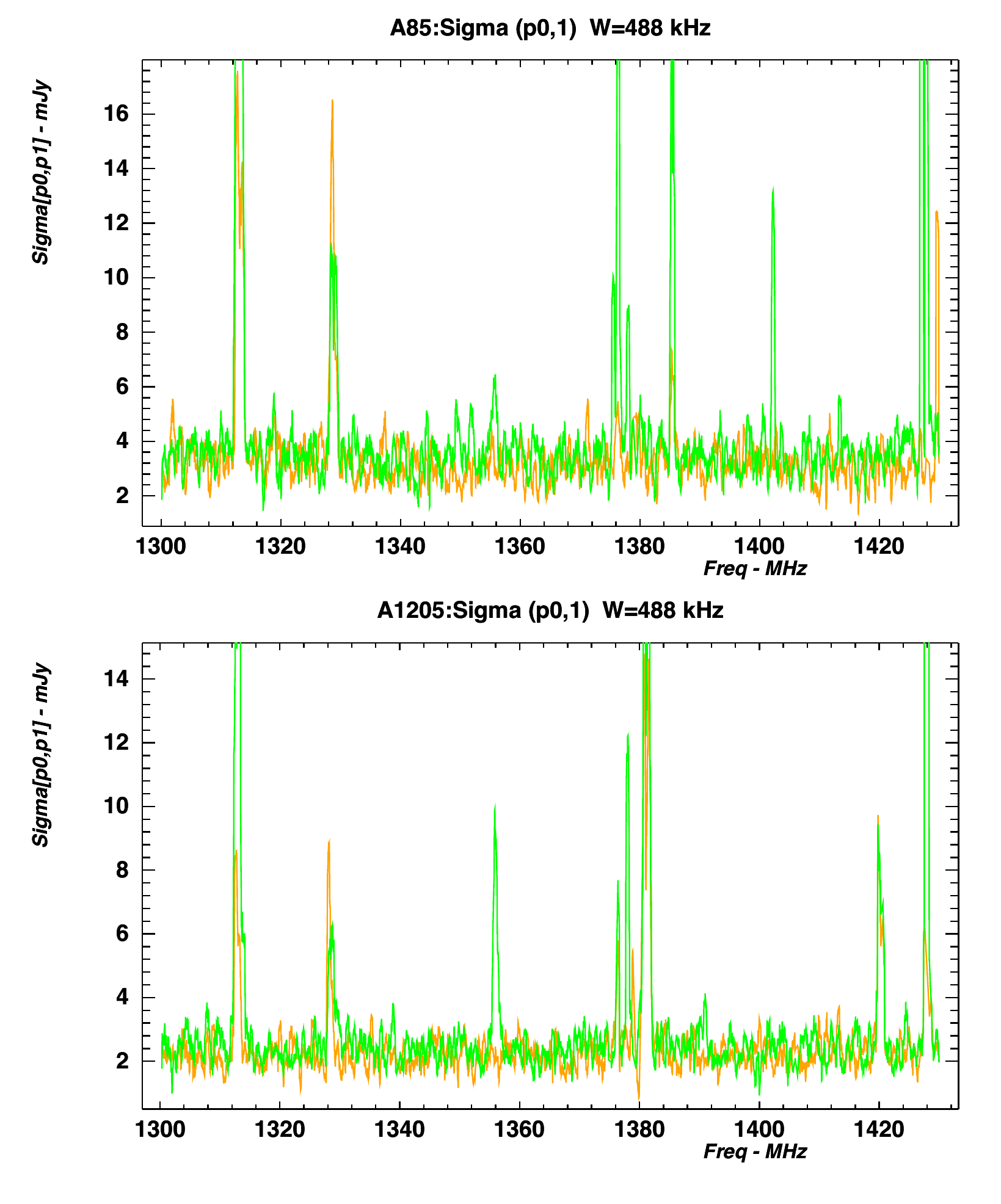}
\caption{Noise level (fluctuations) in mJy as a function of frequency for both polarizations,
estimated from the standard deviation (sigma) along
the frequency,   using a 488~kHz sliding window over the normalized 
$\mathrm{\langle On/g_{on} - Off/g_{off}\rangle}$ 
spectra calibrated using the Milky Way 21 cm emission from the full \Abell{85} and \Abell{1205} data sets. Similar results
have been obtained with the \CalibSrc\, calibrated $\langle\mathrm{(On-Off)/Off}_{filt}\rangle$ spectra
over the full frequency band}
\label{Fig-Abell1205-85-Sigma}
\end{figure}

Figure \ref{Fig-Abell1205-85-Sigma} shows the fluctuation or noise level for the  $\mathrm{\langle On/g_{on} - Off/g_{off}\rangle}$ 
spectra calibrated using the Milky Way 21 cm emission. The noise level has been computed as the standard deviation 
($\sigma$)  along the frequencies, in a $16 \, \mathrm{bins}$ or $\Delta \nu = 488 \mathrm{kHz}$ wide sliding window.
Although the corresponding figure for the \CalibSrc\, calibrated $\langle\mathrm{(On-Off)/Off}_{filt}\rangle$ is not shown, we obtain 
very similar values for the fluctuation level, over the full frequency band. This is an additional indication that 
the calibration procedures described in Sec.~\ref{Sec:Calibration} and Sec.~\ref{Sec:Calibration_3C161} are compatible. 

The noise level is nearly flat for most of the frequency band $[1300, 1430]~\mathrm{MHz}$, with 
single polarization noise level  $\sigma_{1 pol} \lesssim 4 \, \mathrm{mJy}$ for \Abell{85}, and 
$\sigma_{1 pol} \lesssim \, 2.75 \mathrm{mJy}$
for \Abell{1205}. There are however few narrow frequency bands where the fluctuations are significantly higher
around $\nu \simeq 1313$, 1329,  1356, 1376, 1382, 1386, 1402$~\mathrm{MHz}$. 
In the following section, we will use $\sigma_{A85} \simeq 2.7 \mathrm{mJy}$ and 
$\sigma_{A1205} \simeq 1.8 \mathrm{mJy}$ for the signal corresponding to the average of the 
two polarizations $s(\nu) = (p0+p1)/2$, for a single frequency bin ($\simeq 30.5~\mathrm{kHz}$),
for \Abell{85} and \Abell{1205} respectively. This noise level is compatible with the one shown 
in Fig.~\ref{Fig-Abell1205-2440-85-Sensit}. 

The expected noise level in $\mathrm{mJy}$ for the sum of the two polarization signal,
for an instrument characterized by a collecting area $A$ and efficiency $\eta$, 
a system temperature $T_{sys}$, total integration time $t_{int}$ and bandwidth $\delta \nu$ 
can be written as:
$$ \sigma^{2pol} = 10^{29} \times \frac{\sqrt{2} \, k_B \, T_{sys} }{\eta A \, \sqrt{t_{int} \times \delta \nu} } \hspace{5mm} \mathrm{mJy}$$
Assuming a value of the system temperature $T_{sys} \simeq 35~\mathrm{K}$ for NRT, a collecting area $A = 7000 \, \mathrm{m^2}$,
a point source efficiency of $\eta \simeq 0.56$ and a frequency bandwidth $\delta \nu = 30.5 \mathrm{kHz}$ corresponding to the BAORadio spectral resolution, we obtain:
$$ \sigma^{2pol}(t_{int}) \simeq \frac{100}{\sqrt{t_{int}} } \hspace{2mm} \mathrm{mJy} $$ 

The expected noise level would then be $\sigma^{2pol} \simeq 2 \mathrm{mJy} $ for $t_{int} = 2000~s$ integration time in good agreement with measured values from our analysis pipeline by the two calibration methods (Sec.~\ref{Sec:Calibration} and Sec.~\ref{Sec:Calibration_3C161}). Moreover, this result presented here toward the three targets in three different frequency bands suggests that BAORadio system and analysis pipeline is robust over a large frequency domain.

%
%
\subsection{\HI signal search}
\label{sec-hi-signal}

The standard procedure to process ACRT data, in case of known line search, is to make a polynomial baseline fit on $\langle\mathrm{(On-Off)/Off}_{filt}\rangle$ spectra, which allows suppressing the global offset due to gain variations and the oscillation residuals in the spectrum before calculating the fluctuation dispersion. Of course, the line which is searched for in the spectrum is masked to protect it from suppression, hence it is necessary to know precisely both its frequency and width. 

In the analysis presented here we do not have precise information about the \HI signals we are searching for in the clusters. In the \HI intensity mapping this would be even more true as individual galaxy emission would not be significant. Given the NRT beam width, the confusion also increases quickly with redshift. In the case of the observed 
targets here, and as it can be seen in Figs.~\ref{Fig-A85-gals-z} and \ref{Fig-A1205-gals-z}, several possible \HI lines 
might be present in a narrow (few MHz) frequency band. Some of them are emission lines due to galaxies in the On beam while 
others are looking like absorption lines due to galaxies in the OFF beam.

We have searched for  extra power above the noise level around   each cluster redshift frequency (Table~\ref{Tab:Abell}) using data 
from the BAORadio system which is more sensitive than ACRT system. We have performed the cluster 
signal search using the $\mathrm{\langle On-Off\rangle/Off_{filt}\rangle}$ spectra calibrated by the \CalibSrc\, analysis, as well as\\ 
$\mathrm{\langle On/g_{on} - Off/g_{off}\rangle}$ spectra calibrated using the Milky Way \HI line (Sec.~\ref{Sec:Calibration}). 
Both sets of spectra have been Doppler shift corrected. We obtain compatible results for the \HI emission 
from \Abell{85} and \Abell{1205} using both sets of spectra.  For the sake of clarity, we present here only the analysis 
of the  $s(\nu) = \mathrm{\langle On/g_{on} - Off/g_{off}\rangle}$ spectra, as this second method insures that the average signal 
is zero, making the search for emission or absorption lines easier.

\begin{table*}
\small
	\caption{List of frequency peaks above $\pm1\sigma$ in the Sum spectrum of Fig.~\ref{Fig-Signal-Abell85-1205} and the result of investigation of possible origin of these extra power emission during our observation}
	\label{Tab:RFI-sources-Abell85}
         \centering
	\begin{tabular}{cl}
	\tableline
	Frequency & Comments \\
	\tableline
	1332.0~MHz & the peak has been found as a RFI occurring one day\\
	1336.6~MHz & the peak is due to radar tail emission \\
	\multirow{2}*{1350.0~MHz} & at this frequency there is a strong radar emission line\\ 
			   & and      we have also a well identify electronic noise \\
	1353.0~MHz & no human RFI emission found responsible of this peak \\
	1361.3~MHz & the same kind of radar tail emission as 1336.6~Mhz \\
  \tableline
\end{tabular}
\end{table*}

The spectra from the mean of two polarization channels (i.e. $(p_0+p_1)/2$) averaged over a sliding window of size 
$16 \, \mathrm{bins}$ or $\Delta \nu = 488~\mathrm{kHz}$ from \Abell{85} and \Abell{1205} 
data are shown on the figure \ref{Fig-Signal-Abell85-1205} below. 
Frequency components exhibiting large fluctuations as can be seen for instance 
in Fig.~\ref{Fig-Abell1205-85-Sigma} were excluded when computing the average power in the 
$488~\mathrm{kHz}$ sliding window. Notice that the
displayed points are separated by 124~kHz, and the vertical 
scales differ on the right and left hand side plots. 
As expected, the fluctuations are smaller for \Abell{1205} compared to
\Abell{85} given the longer integration time and lower noise level.   
The 40~MHz wide frequency range around the \Abell{85} cluster  $[1330, 1370]~\mathrm{MHz}$ 
is represented on the right hand side of Fig.~\ref{Fig-Signal-Abell85-1205}, while the plots on the left side show 
the 40~MHz frequency band centered on \Abell{1205} $[1300, 1340]~\mathrm{MHz}$. 

The signal we are searching for is unpolarized  while RFI will appear often more strongly in one linear polarization. The difference of the two polarizations signal is also shown on the above mentioned figure (red triangles) and should be 
compatible with zero for an \HI signal.  
We have also represented the +3~sigma detection threshold for 
\Abell{85} and \Abell{1205} based on the noise levels determined in Section \ref{Sec:sensitivity},
taking into account the sliding window size.

We have investigated possible terrestrial origins for extra power emission in the frequency bands 
around \Abell{85}, \Abell{1205} redshifts and for instance the corresponding results for \Abell{85} are summarized in Table~\ref{Tab:RFI-sources-Abell85}.
We have also searched in the frequency range $[1330, 1370]$~MHz for possible RFI emissions responsible 
for the emission in the data taken during \Abell{1205} and \Abell{2440} observations as the three clusters 
have been observed during the same period.

As already mentioned, \Abell{1205} data have a lower noise level, due to a larger total integration time.
Comparing \Abell{85} and \Abell{1205} spectrum around 1350 MHz (Fig.~\ref{Fig-Signal-Abell85-1205}), 
\Abell{1205} spectrum does not show  any extra power around the central frequency, while emissions 
with statistical significance exceeding 3 sigma are observed around  1350 MHz and 1353 MHz in the \Abell{85} spectrum.
The origin of the observed emission around 1350 MHz can not be unambiguously attributed to emission 
from galaxies in the beam. Indeed, we have identified intermittent radar emission at 1350 MHz, and 
although no emission is detected at or near 1350 MHz  in the \Abell{1205} spectrum, we do observe
signs of polarized RFI close to this frequency in the \Abell{2440} spectrum. 
On the other hand, we have not found any suspect sporadic emission in our data 
as well as known RFI emission around 1353 MHz. We consider thus that we have detected \HI emission 
from galaxies belonging to \Abell{85} at 99\% confidence level.
 
We also have searched for extragalactic \HI emission in the \Abell{1205} spectrum, around 1320 MHz, corresponding 
1420~MHz line shifted according to this cluster redshift ($z \simeq 0.077$). Comparing the \Abell{1205} and \Abell{85} data around this frequency (see Fig.~\ref{Fig-Signal-Abell85-1205}), we observe no significant feature in the \Abell{85}, while we observe an excess power in the frequency band $[1315, 1320]~\mathrm{MHz}$, and an even narrower feature 
exceeding marginally the 3-sigma threshold around 1318~MHz. A small frequency band below 1315~MHz were 
subject to RFI and has been blanked. Although the statistical significance is weaker than the \Abell{85}, 
we consider that we have also detected \HI emission from \Abell{1205} at 1318~MHz. 

We do not show the \Abell{2440} spectrum as the corresponding observation time represent less than 1/3 
of the time spent on \Abell{85} and as the frequencies corresponding to \Abell{2440} redshift 
($z \simeq  0.092 , \nu_0 \simeq 1302 \mathrm{MHz}$) are affected by strong RFI (1292~MHz, and 1301~MHz).

Compared to previous \HI surveys with the NRT \citep{2001A&A...368...64V, 2002A&A...389..405P, 2003A&A...408..465M},
our detection threshold and line strength sensitivities have been lowered by a factor $\simeq 5$, from $\simeq 2 \, \mathrm{Jy \, km/s}$
to  $\simeq 0.4 \, \mathrm{Jy \, km/s}$, for comparable, total integration times ($1\div 3$ hours per target). 
One can notice also that the achieved sensitivity limits, in term of line strength $P^*_{21}$ (Table~\ref{Tab:expected-signal}) is comparable to the 
one obtained by the AGES survey toward Abell1367 \citep{2008MNRAS.383.1519C}.

A more in depth analysis of the observed signals toward \Abell{85} and \Abell{1205}, 
and corresponding \HI mass estimates are presented in the next section. 

\begin{figure*}
\centering
\resizebox{\hsize}{!}{
\includegraphics{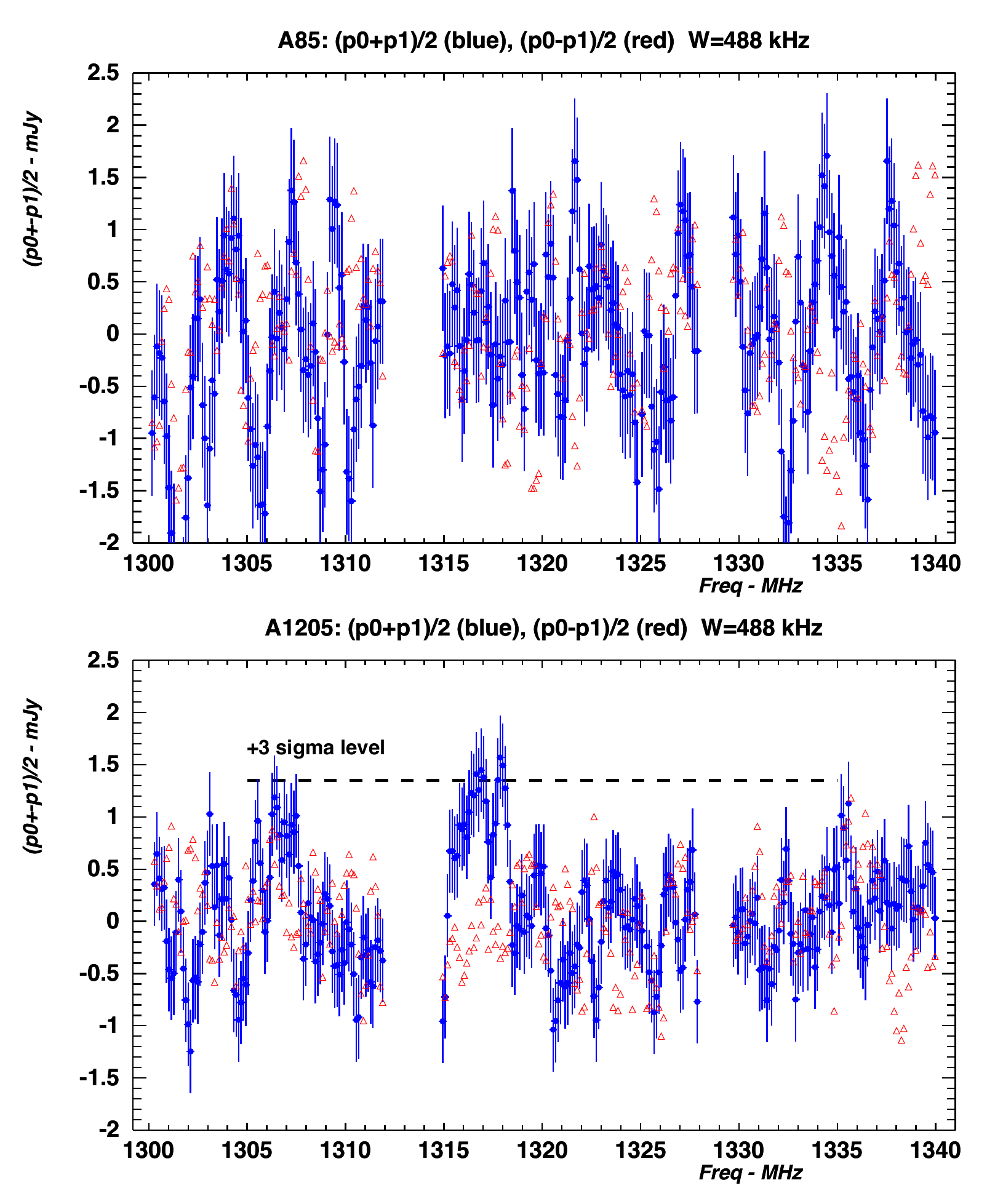}
\hspace{3mm}
\includegraphics{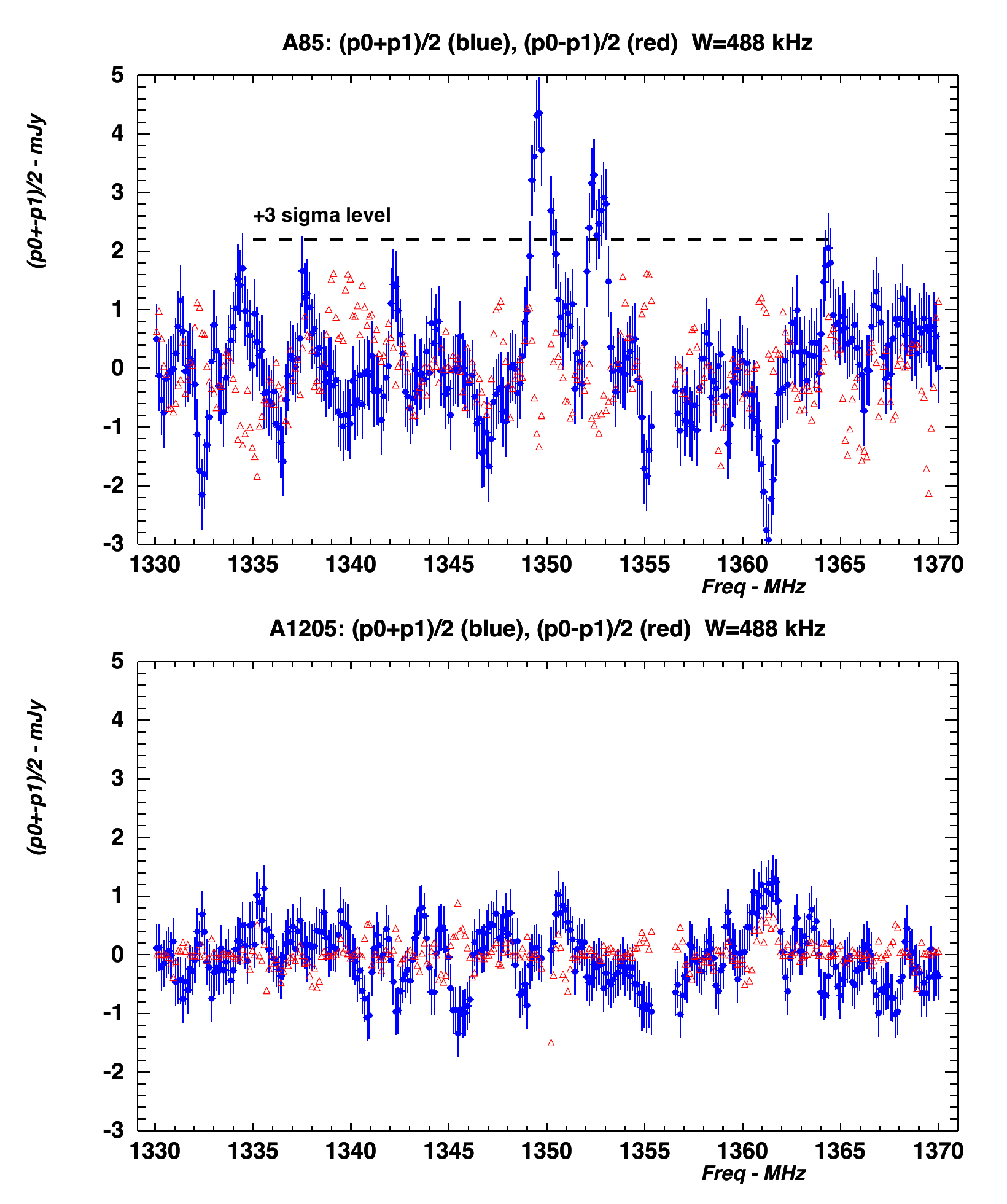} 
}
\caption{Sum (blue circles) and Difference (red triangles) (in mJy) of the two polarization signals
 integrated over a 488~kHz wide sliding window for the \Abell{85} (top) and \Abell{1205} (bottom),
The represented points are separated by 124~kHz. Left: the 40~MHz wide frequency band centered around \Abell{1205} redshift ($z \simeq 0.08, 1320~\mathrm{MHz}$). 
Right: The 40 MHz wide frequency band, centered around \Abell{85} redshift ($z \simeq 0.05 , 1350 \mathrm{MHz}$). 
Notice that the vertical scale is different for the left and right hand set of plots
}
\label{Fig-Signal-Abell85-1205} %
\end{figure*}

\section{Interpretation, discussion}
\label{Sec:InterpDisc}
Using the NED extragalactic database\footnote{NASA/IPAC Extragalactic Database, {\tt http://ned.ipac.caltech.edu/ } } 
and   SDSS DR10 data sets\footnote{ http://www.sdss3.org/dr10/ },
we have identified galaxies in the NRT beam, around the ON source and OFF source
pointing of the telescope for the two targets and in the redshift range ($z \lesssim 0.1$) 
covered by our observations.  The spatial distribution of galaxies, with color coded redshift
around the NRT targets in \Abell{85} and \Abell{1205} are shown in Figs.~\ref{Fig-A85-gals-z} and \ref{Fig-A1205-gals-z}.

\noindent 
\begin{minipage}{0.45\textwidth}
We have also computed the expected signal strength $P_{21}$ from \HI emission in galaxies at the distance 
of observed clusters using the following formula and luminosity distance to the clusters:
$$ P_{21} = 2.4 \, \, 10^{18} \, \frac{M_{\mathrm{H_I}} } {M_\odot } \, \, \mathrm{W} $$ 
The results are summarized in the table shown on the right. 
\end{minipage}
\qquad
\begin{minipage}{0.5\textwidth}
\begin{center}
{ \small
{ Expected total signal strength $P_{21}^*$ received on Earth, from 21 cm emission of a total \HI mass of 
$10^{10}$  \Msol at the three cluster distances studied here. 
$P_{21}^*$ is given in $\mathrm{Jy \, km/s}$ and $\mathrm{mJy.MHz}$ in brackets } \\[1mm]
\begin{tabular}{c c c c c}
\tableline
 & $\nu \, (\mathrm{MHz})$  &  $z$  &  $d_L \, (\mathrm{Mpc})$ & $P_{21}^*$ \\  
\tableline 
\Abell{85} & 1353 & 0.0498  & 228  & 0.830 \, (3.86) \\
\Abell{1205} & 1318 & 0.0777 & 363 & 0.326 \, (1.52) \\
\tableline 
\end{tabular} 
}
\end{center}
\end{minipage} \\[1mm]

The galaxies with redshift measurements have been extracted from the NED database 
around the ON and OFF positions of the NRT beam toward \Abell{85} and \Abell{1205}. Then, these galaxies
were cross matched with SDSS-DR10 galaxies with photometric or spectroscopic redshifts. 
The list of galaxies\footnote{The frequencies listed in Table~\ref{Tab:gallist-A85-A1205} and  the \HI velocities listed in Table~\ref{Tab:fitres-A85-A1205}  have been computed according to the optical astronomer convention and the radio astronomer convention, respectively.} relevant for the study discussed here is presented in Table~\ref{Tab:gallist-A85-A1205}. 
Assuming a gaussian beam shape for the NRT, we have kept galaxies up to a maximum distance 
of 2-sigma from the beam center. We have also applied a selection cut on the redshift, such as the redshifted 
\HI emission frequency falls in the 20 MHz wide frequency band $1342<\nu<1362 \, \mathrm{MHz}$ 
for \Abell{85}, and $1311<\nu<1333 \, \mathrm{MHz}$ for \Abell{1205}. 
We have checked the list of galaxies using the SDSS-DR10 database of galaxies with photometric redshifts,
and we have added one galaxy to the \Abell{85}-On list, which seemed relevant to the study here. 
In total, we have kept 8, respectively 11 galaxies for the \Abell{85}, resp. \Abell{1205}  ON-source position, 
1 and 4 objects for the corresponding OFF-source positions. In addition to object position 
J2000 (RA, DEC), NED magnitude and redshift, we have listed the NRT relative beam efficiency toward the source $\eta_r$, 
assuming a gaussian beam profile $4' \times 22'$, and the SDSS g,r band magnitudes whenever possible.  

The signal spectrum we have obtained toward \Abell{85} and \Abell{1205} have a low signal to noise ratio. 
We have however tried to fit line shapes for multiple sources to our measured spectra. 
The \HI line shape and width depends on the galaxy type, mass and viewing angle. Given our 
low S/N spectra, we have modeled the emission line profiles as gaussians. The non linear fit procedure 
determined the gaussian positions (central frequency) and amplitudes, as well as an overall 
mean signal level assuming line widths. However, approximate best values for these line 
widths has been obtained through an iterative fitting. The FWHM line widths that we have 
obtained range from 380 to 550~kHz (80 - 120~km/s), although weakly constrained. 

\begin{table*}
\centering
\setlength{\tabcolsep}{3pt}
\small
\caption{List of galaxies in the NRT ON/OFF beam observed toward \Abell{85} (first table) and \Abell{1205} (second table), 
extracted from the NED database, identified as galaxies with redshift measurement. In each table, the first list, 
galaxies numbered  with (+) superscript correspond to the galaxies in the ON beam, 
while the second list with (-) superscript, corresponds to galaxies in the OFF Beam. 
{\tt mag}  is the magnitude by the NED database, while {\tt magg, magr} are the SDSS DR10 
magnitudes in the g,r bands, for the associated SDSS source. $\eta_r$ is NRT relative beam efficiency towards the source, assuming 
a gaussian beam with FWHM equal to 4' in RA, 22' in DEC. Sources marked as {\tt Pz} are extracted from 
SDSS DR10 galaxy tables with photometric redshifts  }
\label{Tab:gallist-A85-A1205}
{\small 
\begin{tabular}{llcccccccc}
\tableline
 Galaxy label & Obj. Name & RA (deg) & DEC (deg) & redshift  & mag & $\nu$ & magg & magr & $ \eta_r$ \\
 \tableline
\multicolumn{10}{c}{\Abell{85}} \\ 
 \tableline
G1$^{(+)}$ & 2MASX J00431039-0903243 & 10.7934 & -9.0569 & 0.0578 & 16.40 & 1342.81 & 16.3 & 15.5 & 0.60 \\
G2$^{(+)}$ & SDSS J004310.94-092239.7 & 10.7956 & -9.37772 & 0.0559 & 18.46 & 1345.18 & 18.9 & 18.1 & 0.17 \\
G3$^{(+)}$ & SDSS J004322.80-091635.0 & 10.8450 & -9.27641 & 0.0539 & 18.91 & 1347.72 & 19.3 & 18.6 & 0.42 \\
G4$^{(+)}$ & SDSS J004310.95-091800.3 & 10.7957 & -9.30009 & 0.0532 & 18.62 & 1348.67 & 18.7 & 18.2 & 0.34 \\
G5$^{(+)}$ & SDSS J004319.53-090912.9 & 10.8314 & -9.1536 & 0.0502 & 19.15 & 1352.46 & 19.5 & 19.1 & 0.91 \\
G6$^{(+)}$ & SDSS J004314.35-091021.3 & 10.8098 & -9.17261 & 0.0501 & 18.37 & 1352.65 & 18.9 & 18.6 & 0.87 \\
G7$^{(+)}$ & GALEXASC {\tiny J004315.79-085355.1}   & 10.8160 & -8.89925 & 0.0497 & 19.04 & 1353.17 & - & - & 0.40 \\[1mm]
G8$^{(+)}$ (Pz) & SDSS (Photo-z) &       10.8127 & -9.06866 & $0.051 \pm 0.017$ & - & 1351.40  & 19.5 & 19.4 & 0.92 \\
\tableline
G9$^{(-)}$ & 2MASX J00441835-0902545 & 11.0764 & -9.04854 & 0.0574 & 17.4 & 1343.34 & 17.2 & 16.3 & 0.90 \\
\tableline 
\multicolumn{10}{c}{\Abell{1205}} \\ 
 \tableline
G11$^{(+)}$ & SDSS J111500.29+024756.7 & 168.751 & 2.79908 & 0.0799 & 18.5 & 1315.33 & 18.5 & 17.8 & 0.15 \\
G12$^{(+)}$ & SDSS J111505.42+024835.7 & 168.773 & 2.80994 & 0.0795 & 18.3 & 1315.79 & 18.3 & 17.5 & 0.24 \\
G13$^{(+)}$ & 2MASX J11150889+0235435 & 168.787 & 2.59539 & 0.0794 & 17.9 & 1315.88 & 18.1 & 17.1 & 0.96 \\
G14$^{(+)}$ & SDSS J111505.17+025118.8 & 168.772 & 2.85524 & 0.0793 & 17.7 & 1316.07 & 17.8 & 17.3 & 0.14 \\
G15$^{(+)}$ & 2MASX J11150843+0243335 & 168.785 & 2.72596 & 0.0785 & 16.8 & 1317.06 & 16.8 & 16.2 & 0.54 \\
G16$^{(+)}$ & 2MASX J11151499+0239533 & 168.812 & 2.66472 & 0.0783 & 17.7 & 1317.21 & - & - & 0.49 \\
G17$^{(+)}$ & 2MASX J11151546+0232363 & 168.814 & 2.54348 & 0.0779 & 17.2 & 1317.76 & 17.1 & 16.1 & 0.59 \\
G18$^{(+)}$ & WISEPC J111459.91+024551.0 & 168.750 & 2.76446 & 0.0765 & 17.4 & 1319.45 & 17.5 & 17.4 & 0.19  \\
G19$^{(+)}$ & SDSS J111516.72+022308.7 & 168.820 & 2.38577 & 0.0764 & 18.2 & 1319.63 & 18.3 & 17.5 & 0.28 \\
G20$^{(+)}$ & SDSS J111508.37+023301.4 & 168.785 & 2.55039 & 0.0761 & 18.2 & 1319.95 & 18.2 & 17.8 & 1.00 \\
G21$^{(+)}$ & 2MASX J11151316+0224125 & 168.805 & 2.40360 & 0.0731 & 17.6 & 1323.67 & 17.5 & 16.6 & 0.52 \\
\tableline 
G22$^{(-)}$ & SDSS J111611.93+022237.1 & 169.050 & 2.37700 & 0.0768 & 18.0 & 1319.08 & 18.0 & 17.4 & 0.48 \\
G23$^{(-)}$  & SDSS J111604.50+022946.6 & 169.019 & 2.49630 & 0.0763 & 18.2 & 1319.68 & 18.1 & 17.1 & 0.81 \\
G24$^{(-)}$  & 2MASX J11161766+0221129 & 169.074 & 2.35358 & 0.0763 & 17.3 & 1319.68 & 17.5 & 16.8 & 0.19 \\
G25$^{(-)}$  & 2MASX J11162139+0235209 & 169.089 & 2.58901 & 0.0742 & 18.7 & 1322.25 & 17.2 & 16.2 & 0.17 \\
\tableline 
\end{tabular} 
}
\end{table*}

We have modeled the \Abell{85} spectrum in the frequency range as a sum of $n=5$ gaussian profiles, plus a constant 
in the frequency range $1344-1358 \, \mathrm{MHz}$, while the \Abell{1205} spectrum in the frequency range 
 $1310-1324 \, \mathrm{MHz}$ has been modeled as a sum of $n=9$ gaussian profiles, plus a constant term. It yields
$$ s(\nu) = C + \sum_{i=1}^n \, A_i \, \exp\left( - \frac{(\nu - \nu_i)^2}{2 \sigma_i^2} \right) $$

\begin{figure}[h]
\centering
\includegraphics[width=0.5\textwidth]{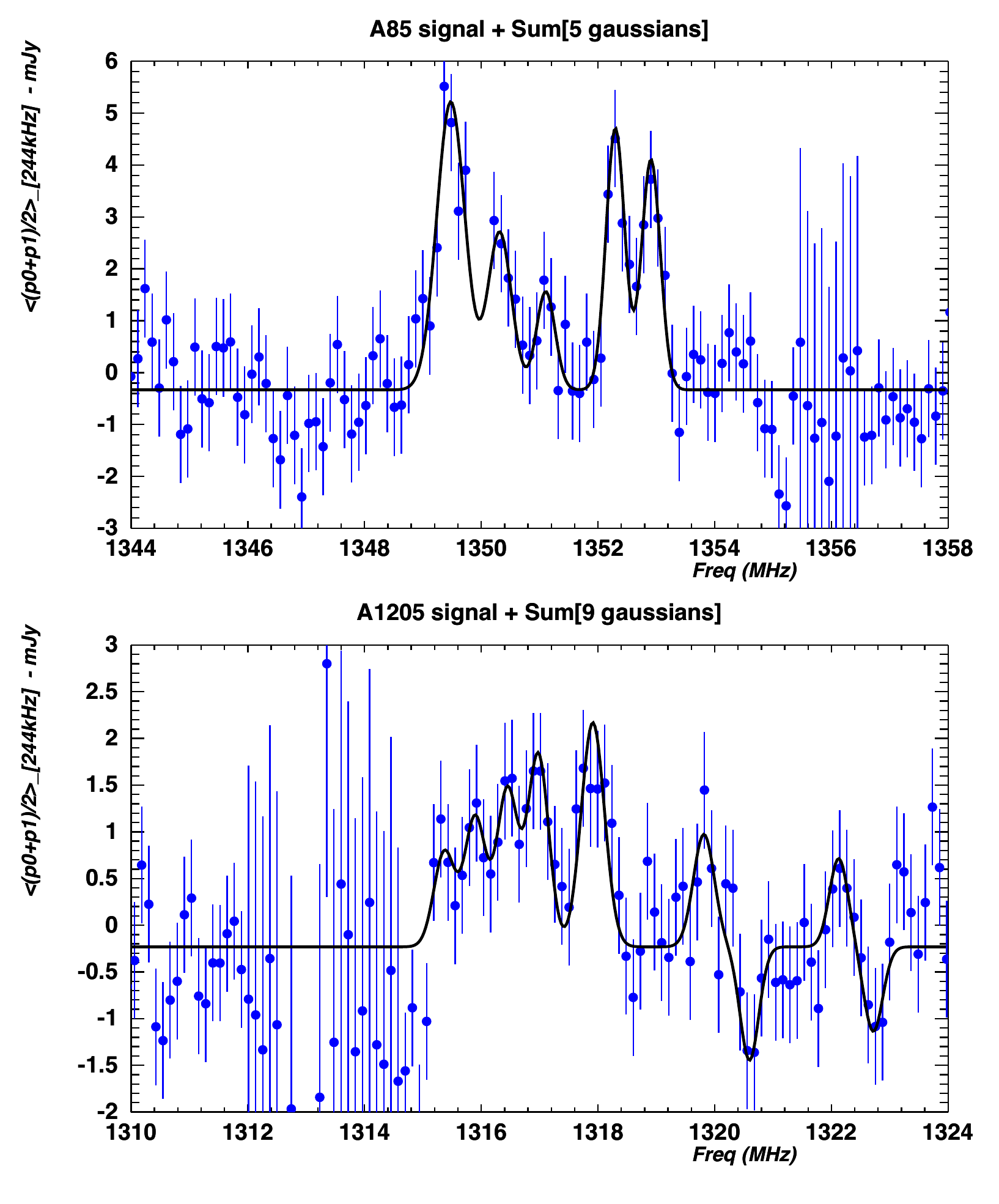}
\caption{Result of the fit of the \HI signal modeled as a sum of 5 gaussian profiles for \Abell{85} (top), in the frequency range 
$[1344, 1358] \mathrm{MHz}$ and as a sum of 9  gaussian profiles for \Abell{1205} (bottom), in the frequency range 
$[1310, 1324] \mathrm{MHz}$. The curve (black line) represents the fit result, and  blue circles the actual measurements,
sum of the two polarization measurements, averaged over 244 kHz (8 frequency bins). We have represented 
one point every 122 kHz (4 frequency bins) and error bars are derived from the fluctuations computed along the frequency axis
(see Fig.~\ref{Fig-Abell1205-85-Sigma})}
\label{Fig-fitHI-A85-1205}%
\end{figure}

The fit results for the \Abell{85} and \Abell{1205} are presented in Table~\ref{Tab:fitres-A85-A1205} and  
shown in Fig.~\ref{Fig-fitHI-A85-1205}. The fit has been performed on the unbinned spectrum,
the fitted model represented as the solid black line, and the measured spectra, averaged over 244 kHz 
wide window represented as blue circles with error bars.  \\[1mm]

\noindent {\bf \Abell{85}:} 
\begin{enumerate}
\item We have no galaxy in our list (see Table~\ref{Tab:gallist-A85-A1205}) that could be associated to the 
two emission lines L1, L2 at 1349.48 and 1350.3 MHz.
As mentioned in the previous section, these two emission like features might be due to imperfectly
cleaned RFI. 
\item The 1.9 mJy line L3 at 1351.1 MHz 
might be associated with the galaxy labeled $G8$ in the \Abell{85} list. 
Assuming $\eta \simeq 0.9$, the estimated \HI mass would be : 
\begin{eqnarray*}
L3: \hspace{1mm} M_{\mathrm{H_I}} \simeq 2.3 \pm 1 \times 10^9 \, M_\odot 
  \hspace{2mm} (V_{HI} = 14625 \pm 20 \, \mathrm{km/s}) 
\end{eqnarray*}
\item The two lines L4, L5 at 1352.3 , 1352.9 MHz are probably associated with the galaxies 
labeled   $G5,G6,G7$. Assuming $\eta \simeq 0.9$ for L4 and $\eta \simeq 0.4$ for L5, 
we  obtain the following \HI mass estimates: 
\begin{eqnarray*}
L4: \hspace{1mm} M_{\mathrm{H_I}}  \simeq  5.6 \pm 1.4 \times 10^9 \, M_\odot 
  \hspace{2mm} (V_{HI} = 14375 \pm 20 \, \mathrm{km/s}) \\
L5: \hspace{1mm}  M_{\mathrm{H_I}}  \simeq  1.1 \pm 0.3 \times 10^{10} \, M_\odot  \hspace{2mm} 
(V_{HI} = 14245 \pm 10 \, \mathrm{km/s})  \end{eqnarray*}
\end{enumerate}

\noindent {\bf \Abell{1205}:} 
\begin{enumerate}
\item The first three lines L11-L13 (at 1315.4. 1315.9, 1316.5 MHz) might be associated  
with galaxies labeled $G11$ to $G14$ in \Abell{1205} galaxy list. The total associated neutral hydrogen mass
would be   $M_{\mathrm{H_I}} \simeq 12. \times 10^{9} \, M_\odot$ , for $\eta \simeq 1$. 
\item The line L14 at 1317 MHz can be associated to the \Abell{1205} galaxies
labeled $G15, G16$, while the line L15 at  1317.9 MHz  is likely to be
due to the G17 galaxy. Assuming $\eta \simeq 0.6$, we obtain the associated neutral
hydrogen masses :
\begin{eqnarray*}
L14: \hspace{1mm} M_{\mathrm{H_I}} \simeq 10 \pm 3 \times 10^{9} \, M_\odot 
  \hspace{2mm} (V_{HI} = 21825 \pm 20 \, \mathrm{km/s}) \\
L15: \hspace{1mm} M_{\mathrm{H_I}} \simeq  13 \pm 3 \times 10^{9} \, M_\odot  \hspace{2mm} 
  \hspace{2mm} (V_{HI} = 21635 \pm 20 \, \mathrm{km/s}) 
\end{eqnarray*}
\item The fitted line L16 at 1319.8 MHz is probably associated with the \Abell{1205} galaxies labeled $G18, G19, G20$,
with a total hydrogen  mass of $M_{\mathrm{H_I}} \simeq 3 \pm 1.5 \times 10^{9} \, M_\odot$ with $\eta \simeq 1$.
\item The other fitted lines (L17, L18, L19) cannot be associated to our knowledge to galaxies in the ON or OFF beam although 
the Table~\ref{Tab:gallist-A85-A1205} might be incomplet.
\end{enumerate}

Integrating simply the \Abell{85} spectrum in the frequency band $\nu \in [1351, 1353]$ will yield a 
total \HI brightness of $P_{21}^* \simeq 0.75 \, \mathrm{Jy \, km/s}$, corresponding to a total mass 
of $M_{\mathrm{H_I}} \simeq 1.5 \, 10^{10} \, M_\odot $ for $\eta \simeq 0.6$,
while performing a similar integration on the \Abell{1205} spectrum
in the frequency band $\nu \in [1315, 1318.5]$ will yield a 
total \HI brightness of $P_{21}^* \simeq 0.65 \, \mathrm{Jy\, km/s}$ corresponding to a total hydrogen mass 
$M_{\mathrm{H_I}}\simeq 3.3 \, 10^{10} \, M_\odot $. These values are in agreement with the ones given above. 
It should also be noted that the quoted uncertainties are underestimated, as they do not include 
line width uncertainty, systematic errors due to residuals in the spectrum shape, calibration 
and beam efficiencies.

\begin{table}[h]
\small
\caption{Result of the multiple gaussian line profiles fit to the ON-OFF spectrum measured toward \Abell{85} (top) and \Abell{1205} (bottom). 
The reduced $\chi^2$ of the fit for \Abell{85} with 5 gaussian line profiles is 0.96 ($N_{dof}=470$), compared to  1.08 for  constant fit.
For \Abell{1205}, the 7 gaussian line profile fit gives a reduced $\chi^2$ of 1.07 ($N_{dof}=505$), compared to 1.15  for a constant fit }
\label{Tab:fitres-A85-A1205}
\centering 
\begin{tabular}{lcccc}
\tableline
Fitted line label & $\nu_0 (\mathrm{MHz})$ & $V_{HI} \, (\mathrm{km/s})$  & $A (\mathrm{mJy})$ & $P_{21}^* \, (\mathrm{Jy \, km/s})$ \\
\tableline
\multicolumn{5}{c}{\Abell{85}}\\
\tableline
L1$^+$ & $1349.48 \pm 0.05$ &  $14 970 \pm 10 $ & $5.6 \pm 0.8$ & $0.70 \pm 0.09$ \\  
L2$^+$ & $1350.30 \pm 0.10$     &  $14 795 \pm 20 $ & $3.0 \pm 0.9$ & $0.32 \pm 0.09$ \\  
L3$^+$ & $1351.10 \pm 0.10$     &  $14 625 \pm 20 $ & $1.9 \pm 0.9$ & $0.17 \pm 0.08$ \\  
L4$^+$ & $1352.30 \pm 0.10$     &  $14 375 \pm 20 $ & $5.0 \pm 0.9$ & $0.42 \pm 0.08$ \\
L5$^+$ & $1352.90 \pm 0.05$   &  $14 245 \pm 10 $ & $4.4 \pm 0.9$ & $0.38 \pm 0.08$ \\
\tableline 
\multicolumn{5}{c}{\Abell{1205}}\\
\tableline
L11$^+$ & $1315.4 \pm 0.15$ &  $22 160 \pm 30$ &  $1.0 \pm 0.6$    & $0.10 \pm 0.06$ \\  
L12$^+$ & $1315.9 \pm 0.13$ &  $22 055 \pm 30$ &  $1.4 \pm 0.6$  & $0.13 \pm 0.06$ \\  
L13$^+$ & $1316.5 \pm 0.10$   &  $21 930 \pm 20$ &  $1.7. \pm 0.6$ & $0.16 \pm 0.06$ \\  
L14$^+$ & $1317.0 \pm 0.08$ &  $21 825 \pm 20$ &  $2.0 \pm 0.5$    & $0.19 \pm 0.06$ \\  
L15$^+$ & $1317.9 \pm 0.06$ &  $21 635 \pm 15$ &  $2.4 \pm 0.5$  & $0.25 \pm 0.06$ \\  
L16$^+$ & $1319.8 \pm 0.10$   &  $21 230 \pm 20$ &  $1.2 \pm 0.6$  & $0.11 \pm 0.06$ \\
L17$^+$ & $1322.1 \pm 0.15$ &  $20 750 \pm 30$ &  $0.9 \pm 0.6$ & $0.08  \pm 0.05$ \\[1mm]
L18$^-$ & $1320.6 \pm 0.10$   &  $21 065 \pm  20$ &  $1.2 \pm 0.6$ & $0.11 \pm 0.05$ \\
L19$^-$ & $1322.7 \pm 0.15$ &  $20 620 \pm 30$  &  $0.9 \pm 0.6$ & $0.08 \pm 0.04$  \\
\tableline 
\end{tabular}
\end{table}


\section{Conclusions}
\label{Conclusion}

We have investigated the capabilities of the new BAORadio analog and digital back-end, associated with our data acquisition and processing pipeline, through a pilot observation program of search for \HI emission from galaxies in clusters at redshift 
$z \sim 0.05 - 0.1$.   
More than 50 hours of observations have been carried out toward the three clusters (\Abell{85}, \Abell{1205}, \Abell{2440}), and calibration 
sources with the BAORadio system, in parallel with the standard NRT correlator (ACRT), over a period of one year. 
We have shown the superior RFI cleaning performance achieved thanks to the BAORadio electronic 
chain, offering  full digitization at 500~MHz with fine time sampling ($\sim 0.1 \mathrm{ms}$). 

We have also demonstrated the high level of stability of the new system, far better than the standard NRT correlator,
which suffers from the analog signal transmission over several hundred meters.
Surprisingly, we have found that the standard calibration procedure used at NRT, based on noise diode pulses injected 
at the beginning of each observation cycle, is the source of additional signal fluctuations. 
We have thus relied on the system stability, associated with calibration with respect to astrophysical sources,
\CalibSrc\, radio source and the Milky Way \HI emission for the results which have been presented.

We have obtained the radiometer curves for both systems (BAORadio, ACRT) and the three frequency bands centered 
on the Abell cluster redshifts. After about 2000~s of integration time the BAORadio system has reached a sensitivity 
of about 1.4~mJy, while the standard NRT ACRT system sensitivity is at least 5 times worse, even after extensive data cleaning. 
We have also identified an additional noise contribution for integration time larger that 1 s, using the BAORadio 
data, which might be due to the cryogenic cooling system. 

Unfortunately, the spectra that we have obtained presents some structuring as 
a function of frequency. The source of these modulations have been clearly identified: impedance mismatch between the  NRT 
cryogenic amplifier and BAORadio analog board is responsible for the few MHz modulations, while the $\sim500~\mathrm{kHz}$ modulations comes from the standing waves between the NRT spherical mirror and the receiver horns.
It should be noted that $\sim500~\mathrm{kHz}$ modulations can be partially canceled thank to the horizontal motion 
of the NRT focal plane assembly along the direction of the waves propagation, but this feature was disabled during our observations due to mechanical maintenance.
We have used this horizontal motion in subsequent observations which indeed decreases the $\sim500~\mathrm{kHz}$ 
modulation amplitude drastically. 


Although quite challenging given the NRT sensitivity and RFI environment, our search for \HI emission from 
galaxies within clusters in the redshift range $z \lesssim 0.1$ has been successful,
thanks to the BAORadio electronic system performance and our dedicated data reduction pipeline.  We are fairly confident on a detection of emission in \Abell{85}, with $>3 \sigma$ 
significance level, leading to a total \HI brightness of about $0.8~\mathrm{Jy\ km/s}$ in the $[1351, 1353]$~MHz band.
Concerning the \Abell{1205} cluster, we report the detection of a 21 cm emission signal in the frequency band $[1315, 1318]$~MHz,
but at a lower statistical significance. 
The corresponding \HI brightness in the integrated spectrum is about $0.6~\mathrm{Jy\ km/s}$. 
We have performed a cross identification of the detected emission lines with optically detected galaxies and have derived 
mass estimates for galaxies in \Abell{85} and \Abell{1205}.

Obviously, larger instruments such as the Arecibo radio telescope or the upcoming SKA instruments will be more 
effective for such non local \HI searches. It would however be possible to carry more ambitious search 
with NRT and BAORadio using several hundred observation hours. The analog input stage needs however to 
be modified to decrease the impedance mismatch, and the acquisition system has to be upgraded. 
The ON-sky observation efficiency could then easily be pushed to more than 50\%, as we have already demonstrated 
it in test observations. Moreover, most of the CPU and I/O intensive steps of the RFI cleaning 
and data reduction could then be performed on the acquisition computers, easing 
the subsequent data analysis task. 


\begin{acknowledgements}
The observations at Nan\c cay would not have been possible with-out the help and support of the operators and of the technical staff
of the radio telescope.
The Nan\c cay Radio Observatory is the Unit\'e scientifique de
Nan\c cay of the Observatoire de Paris, associated as Unit\'e de Service
et de Recherche (USR) No. B704 to the French Centre National de la
Recherche Scientifique (CNRS). The Nan\c cay Observatory also gratefully acknowledges the financial support of the Conseil r\'egional of
the R\' egion Centre in France. We acknowledge financial support from "Programme National de Cosmologie and Galaxies" (PNCG) of CNRS/INSU, France.

We have made use of the NASA/IPAC Extragalactic Database (NED) which is operated by the Jet Propulsion Laboratory, 
California Institute of Technology, under contract with the National Aeronautics and Space Administration. 
Photometric and spectroscopic data from SDSS has also been used. 
Funding for SDSS-III has been provided by the Alfred P. Sloan Foundation.

We thank Eric Gerard for useful discussions and suggestions. We thank also  J. Pezzani and C. Viou,
D. Charlet, C. Pailler and M. Taurigna for their help and assistance during observations 
and understanding the BAORadio system.  
\end{acknowledgements}


\appendix\normalsize

%

\section{ BAORadio data reduction pipeline details}
\label{App:procpipeline}
\subsection{RFI filtering}
\label{App:RFIFilter}
In the part of the L-band where the local and redshifted \HI line can be search for, only the [1400,1427]~MHz is reserved with a primary status for radio astronomy, so it is a main concern to detect and filter as much as possible the RFI emission in the full surveyed band $[1250, 1500]$MHz and especially the band used for \HI signal in the three observed galaxy clusters. 

\begin{figure}[htb]
\centering
\includegraphics[width=0.47\textwidth]{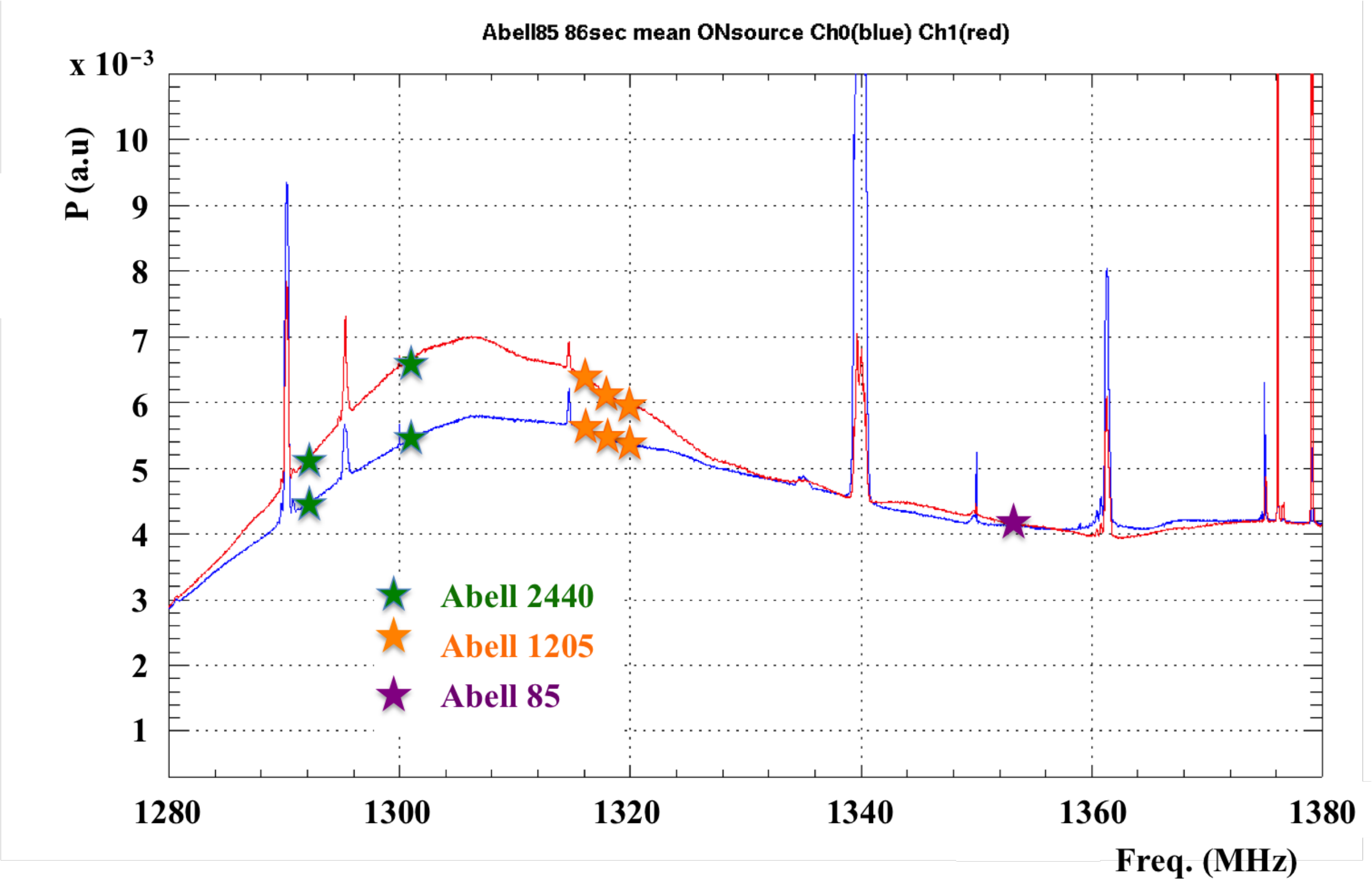}
\qquad
\includegraphics[width=0.47\textwidth]{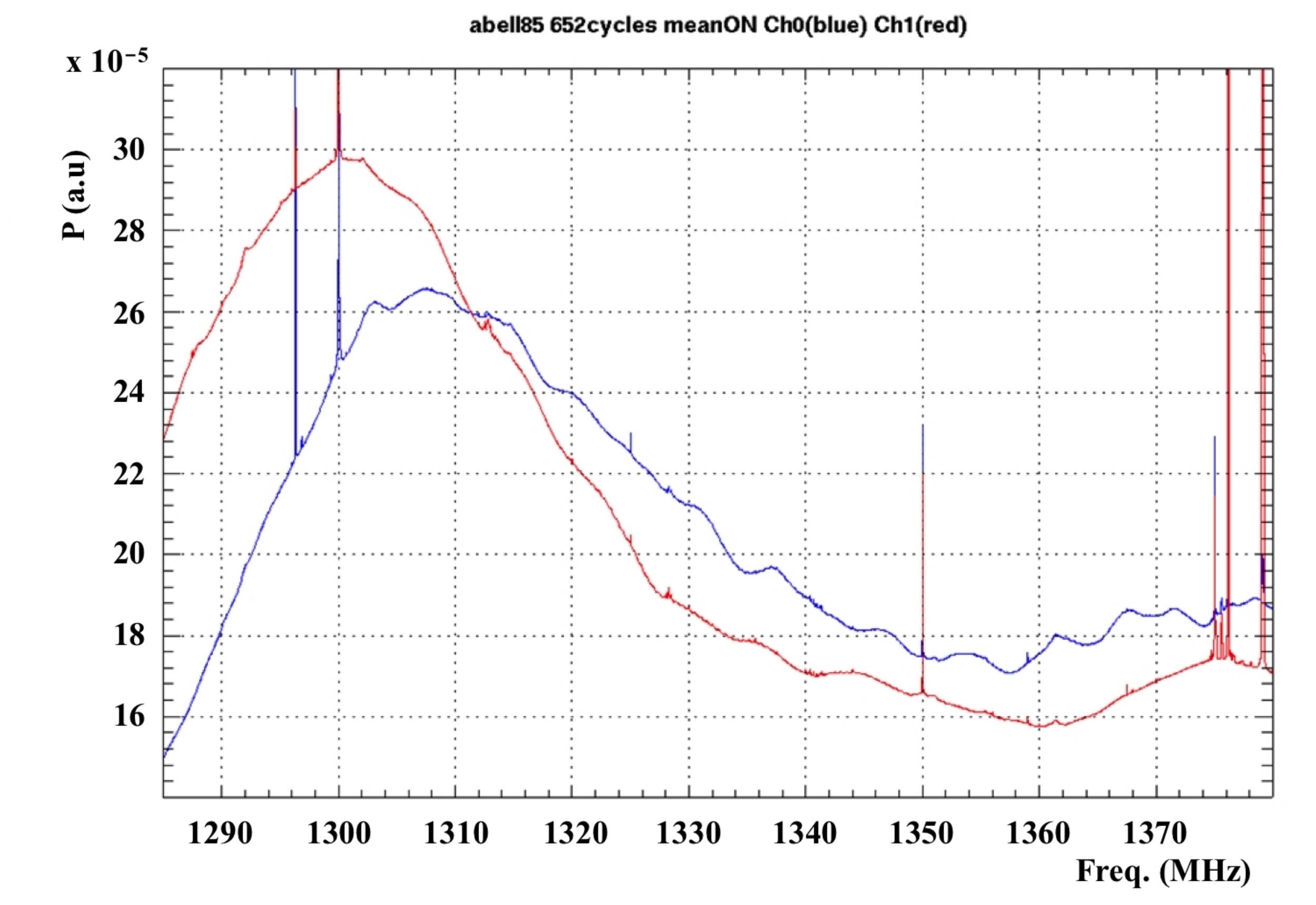}
\caption{left: mean of BAORadio ON spectra without RFI filtering for both polarizations (polar.~0: blue and polar.~1: red), 
  corresponding to a time slice of 86.4~s from \Abell{85} observation; 
  the stars mark the frequencies where the \HI signal is expected (see Table~\ref{Tab:Abell}). 
  right: mean of BAORadio ON-source spectra over the full data set for \Abell{85} observations (i.e. 652 cycles) 
  processed with median filter in time }
\label{Fig-filtrage-median-BAORadio}
\end{figure}

Figure~\ref{Fig-filtrage-median-BAORadio} (top) presents an example of strong RFI signals registered during the \Abell{85} observation by the BAORadio DAQ during 86.4 seconds integration time (3 On cycles). The stars mark the frequency where the \HI signal is expected
for the three clusters. 


With BAORadio, it is possible for instance to produce time-frequency maps with ($0.6$~sec, $30.5$~kHz) resolution by taking the mean of 5120 BRPaquets. Using the same data of Fig.~\ref{Fig-filtrage-median-BAORadio} (top), two such maps are shown in Fig.~\ref{Fig-BAOelec-RFI-1}.

We have identified several RFIs: three strong ones at 1290~MHz, 1340~MHz and 1361~MHz, and four weaker at 1295~MHz, 1315~MHz, 1335~MHz and 1350~MHz, all with cadences of either 5~sec or 10~sec. These signals might be emitted by the civil air traffic control radars in the Parisian region entering by low level backward lobes created by the East-West geometry of the large mirrors.
We find as well electronic RFI at 1300~MHz and 1350~MHz which are permanent, so in the time-frequency image appear as a continuous thin line of thickness given by the lowest possible $\Delta\nu$ of 30~kHz. Such electronic pollutions have been investigated afterwards on test bench and it appears that the most likely source was certainly our local oscillator. 

To tackle the above intermittent RFI, a median filter algorithm has been applied on the 8192 
frequency components along the time axis with a depth of 5120 digitization frame, corresponding to ~0.64 second.

The result obtained is presented in Fig.~\ref{Fig-filtrage-median-BAORadio} (bottom) with all the statistics available (i.e. not only the few seconds of Fig.~\ref{Fig-filtrage-median-BAORadio} (top), showing the gain in terms of removing spikes and decreasing the fluctuations. Our RFI cleaning filter removes most of the intermittent RFI signals listed above. At 1300~MHz and 1350~MHz, the permanent narrow RFIs generated by the electronic system itself, obviously remain after median filtering over time. We also see a residual RFI at about 1296.4~MHz which might be a tail of the radar at 1295~MHz.

\subsection{BAORadio data quality and system stability}
\label{App:BAORadiodataquality}
We have collected data from the three Abell clusters observations from March to December 2011,
using the BAORadio electronic and acquisition system. Figure~\ref{Fig-stability-BAO} shows the average power as function of the integration time in a 1~MHz wide band
around the central frequency $\nu_{obs}$ for each cluster, computed from $\langle\mathrm{(On-Off)/Off}_{filt}\rangle$ spectra, 
in black for the complete data set, and in red, for the selected cycles. 


We have used few selection criteria to reject low quality or noisy data.
The system was unstable during september 2011, although we have not identified clearly the origin of the instability.
The corresponding data sets have been dismissed, as well as observations affected by the Sun transit. 
Finally, we have applied loose cuts on $\langle\mathrm{(On-Off)/Off}_{filt}\rangle$ distribution to reject abnormal observations. 
The selected data sets are shown in red in Fig.~\ref{Fig-stability-BAO} as a function of the cycle number for the three 
clusters. This corresponds to an overall selection efficiency greater than 99\%. 
We remind however that the BAORadio data acquisition useful time fraction ON-sky was 25\% due to network data flow limitation 
(see Sec.\ref{BAORadio}). In addition, \Abell{2440} observations were not used for \HI emission line search due to limited integration 
time and the presence of strong RFI in the frequency range of interest. 


\subsection{Calibration cross-check with \NGC{4383}} 
\begin{figure}
\centering
\begin{minipage}{0.47\textwidth}
	\includegraphics[width=0.95\textwidth]{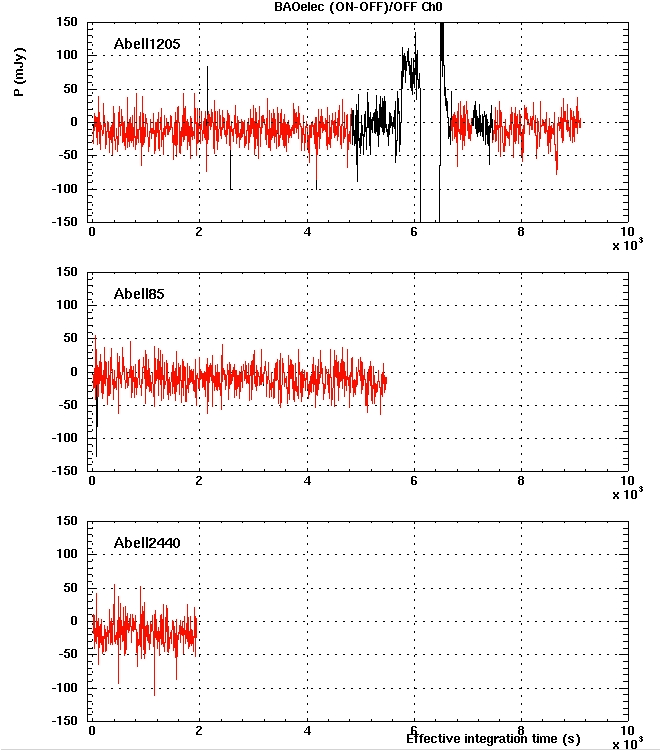}
\caption{Time evolution of BAORadio $\langle\mathrm{(ON-OFF)/OFF}_{filt}\rangle$ integrated over 1~MHz centered on 1321~MHz for \Abell{1205} (top), 1346~MHz for \Abell{85} (middle) and 1302~MHz for \Abell{2440} (bottom). The effective cycle time sampling of  about 8.4~s per data point is used. The whole data set is plotted in black, the data selected for analysis are in red (see text)}
\label{Fig-stability-BAO}
\end{minipage}
\qquad
\begin{minipage}{0.47\textwidth}
   \includegraphics[width=0.95\textwidth]{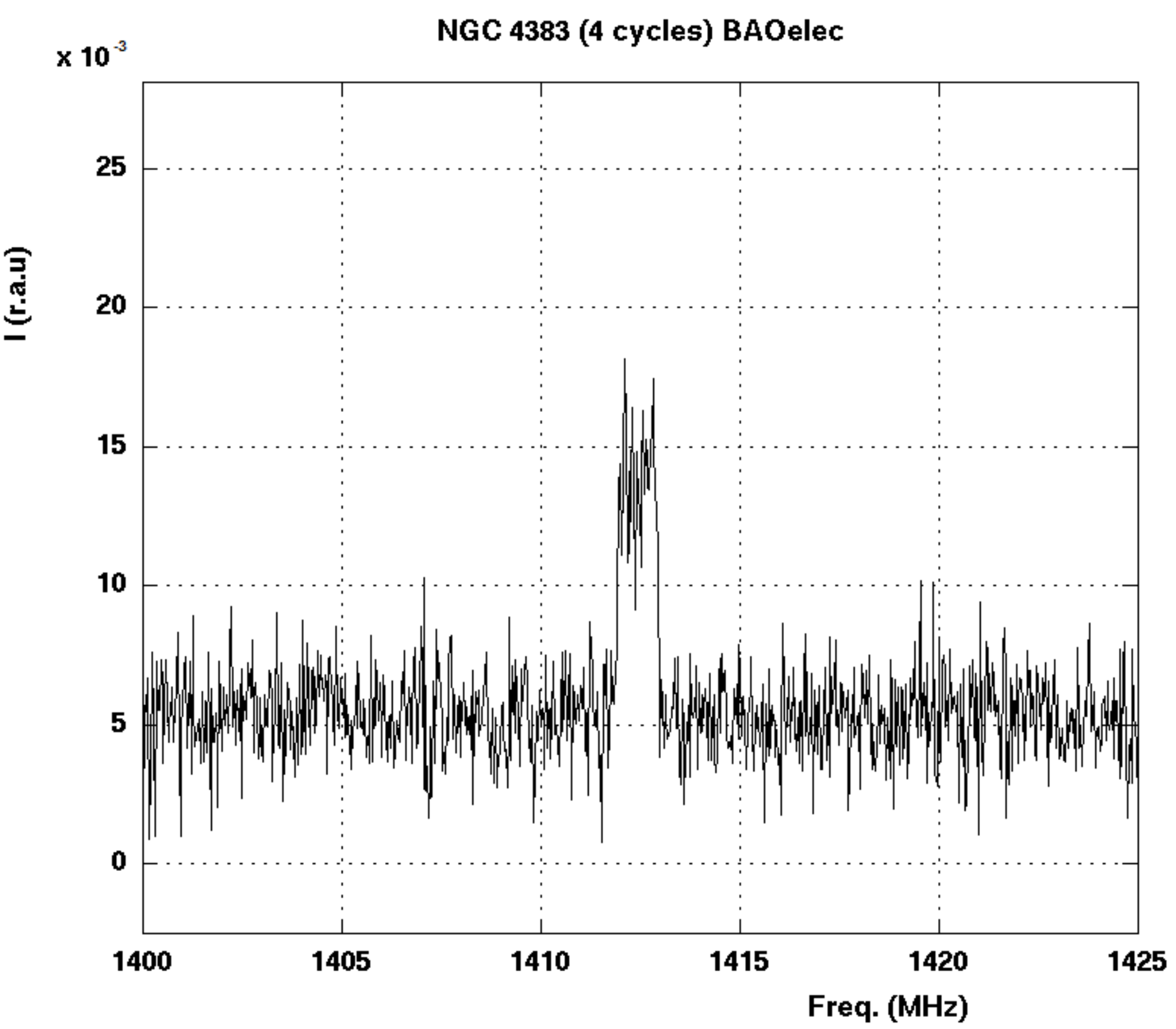}
   \caption{The mean value of $\langle\mathrm{(On-Off)/Off}_{filt}\rangle$ in relative arbitrary units over the two polarizations and over 4 ON-OFF cycles with the BAORadio system on \NGC{4383} source. We see the \HI line well above the continuum }
   \label{Fig-NGC4383-BAORadio}%
\end{minipage}
\end{figure}

To confirm that the \CalibSrc\, calibration method was correct (Sec.~\ref{Sec:Calibration}), we have observed the \NGC{4383}\, source which has both a continuum and a \HI line of $\sim$1-MHz width in the range $[1412, 1413]$~MHz. Figure~\ref{Fig-NGC4383-BAORadio} displays the mean over 4 cycles of $\langle\mathrm{(On-Off)/Off}_{filt}\rangle$ spectrum in the range $[1400, 1425]$~MHz with the BAORadio system. The \HI line is clearly visible well above the continuum and the stochastic noise. Using the calibration coefficient determined with the \CalibSrc\, radiometric source, one finds a \HI line strength of $41.1 \pm 1.4 (stat) \pm 1.1 (syst)$~Jy.km/s which is quite in agreement with the value $48.4\pm5.1$~Jy.km/s reported in reference  \citep{2009AJ....138.1741C,2010AJ....139.2716C}. Notice also that this source is not supposed to be stable and strong enough for calibration purpose.   

\section{NRT standard auto-correlator data analysis} 
\label{App:ACRT}

In this section we describe the analysis done on the data sets from the same observations,
acquired in parallel with the NRT standard auto-correlator (ACRT), in order to compare its performance to the BAORadio system, especially in terms of sensitivity. For this analysis we have used the standard NRT software tools
as well as a specific data reduction pipeline  similar to the BAORadio one.

\subsection{ACRT description} \label{App:ACRTDescription}
The NRT standard auto-correlator consists roughly of a mid-frequency amplifier and a Control Auto de Gain (CAG) that scales the input to 2 V, after which it is split into two parallel chains:
\begin{itemize}
	\item The filter chain measures the total power received in the whole bandwidth vs. time which is digitized with a 12-bit ADC and a SEFRAM  (rapid digitization at 4 to 500~Hz). 
  \item The digital correlator chain (ACRT) measures the total power received and the signal vs. time and computes the auto-correlation of the signal. The output at hardware level is the total power received and the power spectrum normalized to the total power received, both digitized with a 3-bits  9-level ADC.
\end{itemize}
The  digital auto-correlator has a total bandwidth that can be set from 200 kHz to 50 MHz. The 8192 frequency channels can be split into 2 to 8 banks. In our case we have set up a 25-MHz bandwidth and 4 banks which correspond to 2 linear polarizations and 2 circular polarizations (not analyzed). This gives 2048 channels per bank and a frequency resolution of 12.2 kHz. The NRT acquisition chains are 100\% efficient ON-sky, and the minimum integration time is 1-sec. This has to be compared to BAORadio system which had 25\% on sky efficiency, but a  time resolution of 16.7~ms. The NRT provides standard tools (NAPS) to pre-process and calibrate raw observations.



\subsection{ACRT data quality and system stability} \label{App:ACRTdataquality}
%

\begin{figure}
\centering
\includegraphics[width=0.75\textwidth]{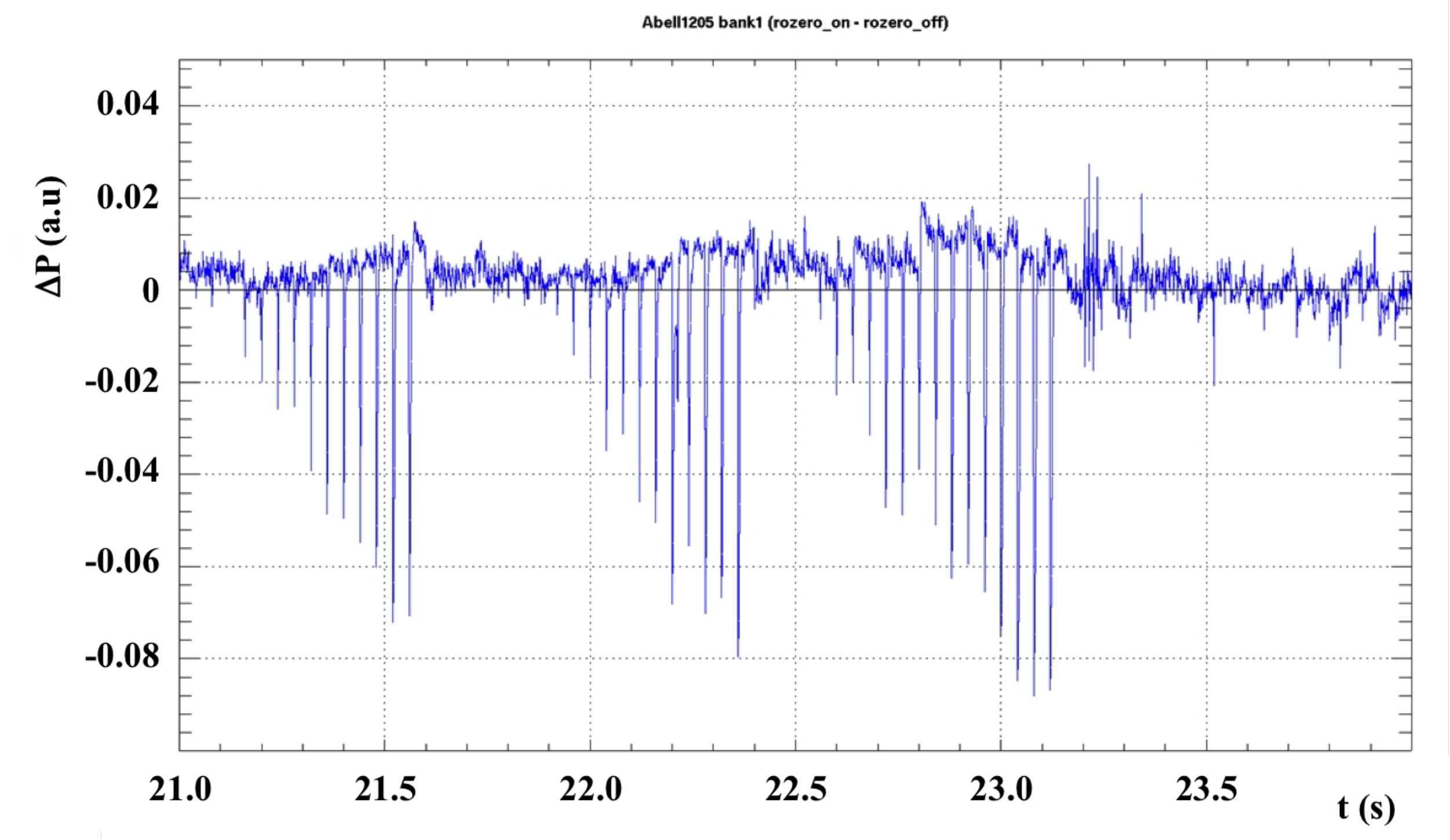}
\caption{Difference of total power received in ON and OFF phases vs. time for polarization p0 to illustrate the effect of cable stretching in the signal. The zero value is marked with a black line}
\label{Fig-pan-flute}%
\end{figure}

The whole ACRT data set extend by one month (i.e. January 2012) the data set acquired with the BAORadio system. 
Similar selection criteria as the ones listed in Section~\ref{App:BAORadiodataquality} were applied, as well as an additional 
cut on the first 10-sec of each cycle as the signal is affected by the stretching of the 150~m long cable carrying 
the analog signal out of the chariot. 
This happens when the chariot repositions itself between the end of the ON phase and the beginning of the OFF phase.
A striking illustration of the effect of the cable stretching can be seen on Fig.~\ref{Fig-pan-flute} where the 
difference of the total power collected in ON and OFF phases for the polarization p0 is presented as function of time. 
Notice the same trend is also observed on p1 (not shown). 
We observe a "pan flute" pattern due to the tail after the DAB pulse during the OFF phase which can last for about 10~sec maximum. This cable stretching is responsible for 25\% loss of 1-sec integrations for ACRT and is the main source of inefficiency. The BAORadio system is not affected by cable stretching as the digitization is done in the chariot as close as possible to the first preamplifier and filters chain.

Similarly to Figure~\ref{Fig-stability-BAO} for BAORadio, we show in Figure~\ref{Fig-stability-ACRT} the time evolution of the mean values $\langle\mathrm{(On-Off)/Off}_{filt}\rangle$ over 1~MHz centered on the central frequency in ACRT observation band for each cluster (see Table~\ref{Tab:Abell}). The whole data set is presented in black and the selected data for analysis (including RFI-cleaning) in red.


Table~\ref{Tab:UseFulData} summarizes the observations kept after the quality cuts for both BAORadio and ACRT. It is clear that ACRT system currently suffers from analog signal propagation, making it less efficient than BAORadio despite its intrinsic ON-sky efficiency of 100\%.
  \begin{table*}
  \small
      \caption{Efficiency and integration time ON-sky of ACRT and BAORadio in terms of fraction of useful cycles used after quality cuts and RFI-cleaning procedure. We remind that the efficiency  ON-sky is 100\% for ACRT and 25\% for BAORadio}
         \label{Tab:UseFulData}
         \centering
         \begin{tabular}{ccrrrr}
            \tableline
            Cluster      &  System & Total cycles & Useful cycles & Efficiency & Integration ON-sky\\
            \tableline
            \multirow{2}{*}{\Abell{1205}} &  ACRT     & 1374      & 741            & 53.9~\% & 29 640~s\\
                       &  BAORadio  & 840       & 837            & 99.6~\% & 6 277~s\\
            \multirow{2}{*}{\Abell{2440}} &  ACRT     & 320       & 141            & 44.1~\% & 5 640~s\\
                       &  BAORadio  & 230       & 230            & 100.0~\% & 1 725~s\\
            \multirow{2}{*}{\Abell{85}}   &  ACRT     & 737       & 287            & 38.9~\% & 11 480~s\\
                       &  BAORadio  & 652       & 651            & 99.8~\% & 4 882~s\\
            \tableline
         \end{tabular}
   \end{table*}

\subsection{ACRT data analysis }

The ACRT system outputs for each polarization the total power received and the autocorrelation of the signal normalized to this power. For the analysis we have taken as "raw spectra" the product of the normalized autocorrelation signal and the total power received for each 1-sec integration. We have then applied a similar analysis method as for BAORadio: selection of data (quality cuts and RFI-cleaning, described below), obtention of $\langle\mathrm{(On-Off)/Off}_{filt}\rangle$ spectrum and time series analysis of the mean signal integrated on 1~MHz band around the cluster observation frequency to obtain the radiometric curve.

We have cleaned the data from RFI using two algorithms. The first one consists of eliminating the integrations that significantly differ in spectral power from the mean of a given phase (ON or OFF), which could be the case of spectra with extra noise or affected by RFI. This is done computing along the frequency axis the standard deviation between the 1-sec integration spectra and a mean frequency spectrum over the whole phase considered, and rejecting those integrations with a standard deviation above a threshold. This algorithm is inspired on the Integration Limit RMS  algorithm, nicknamed "ILR" and implemented on the NRT standard software analysis NAPS \citep{NAPSUserGuide}.
The second RFI-cleaning algorithm searches for intermittent radar signals appearing as peaks in the time evolution of the total power received. As this power is digitized with a small dynamic range, we enhance the peaks considering instead the "total radar power", defined as the sum of 2-MHz bands centered on the known radar signals seen in Fig.~\ref{Fig-BAOelec-RFI-1}.
%

The combination of the two RFI-cleaning algorithms rejects 27\% of the data for \Abell{1205}, 47\% for \Abell{85} and 41\% for \Abell{2440}.

\subsection{ACRT calibration}
\label{App:ACRTCalibration}
During a cycle, diode pulses (DAB) are injected in the NRT horn at the beginning of an acquisition phase (ON and OFF). In general this kind of known injection power is used as default calibration thanks to the conversion relation determined with dedicated observations of well-known calibrator sources during NRT special runs \citep{1990A&AS...86..473F}. But during our cluster observations this routine procedure has shown surprisingly that the system is much more stable than the DAB itself.

\begin{figure}
\centering
\begin{minipage}{0.47\textwidth}
\begin{center}
	\includegraphics[width=0.95\textwidth]{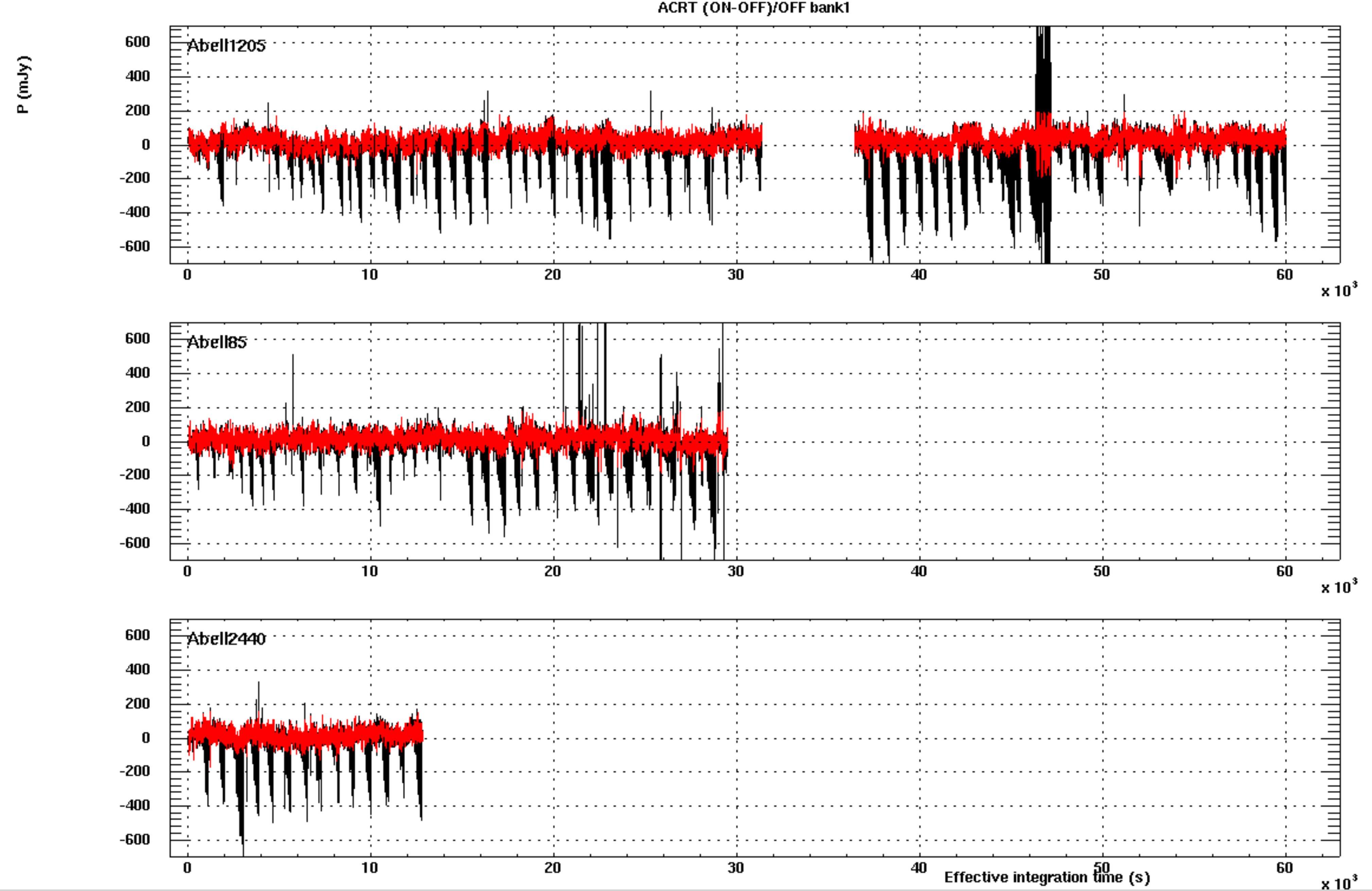}
	\caption{Time evolution of ACRT $\langle\mathrm{(On-Off)/Off}_{filt}\rangle$ integrated signal in 1~MHz band around the observation frequency of each cluster: in black the complete data set of observations, in red the subset of observations kept after quality cuts}
\label{Fig-stability-ACRT}
\end{center}
\end{minipage}
\qquad
\begin{minipage}{0.47\textwidth}
\includegraphics[width=0.95\textwidth]{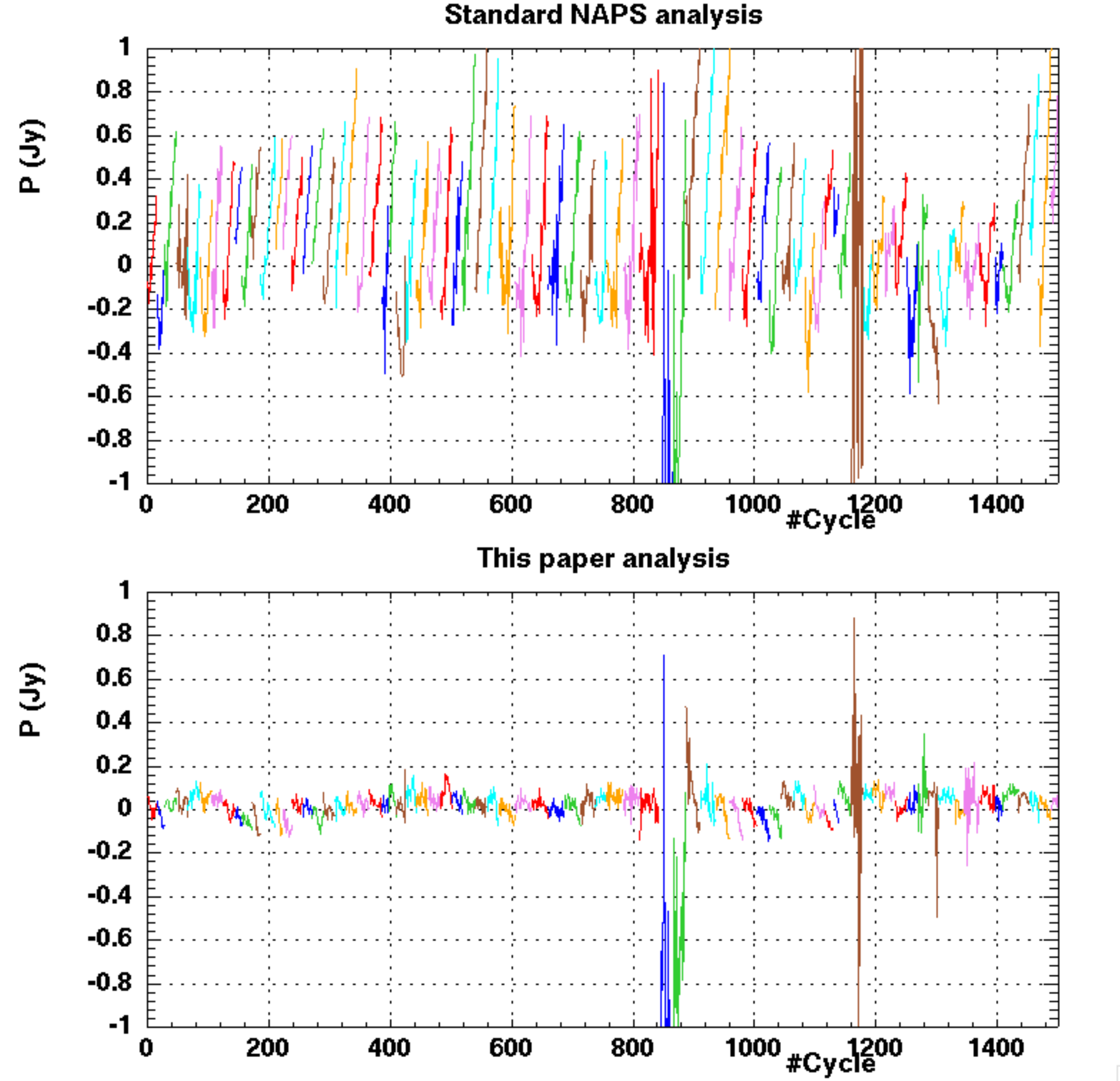}
\caption{ACRT mean $\langle\mathrm{(On-Off)/Off}_{filt}\rangle$ signal integrated in the band $[1320.3, 1321.3]$~MHz versus 
the cycle number for polarization p0, obtained from the standard NAPS analysis (top) and from the analysis presented in this paper (Sec.\ref{Sec:Calibration}) (bottom)}
\label{Fig:NAPS}
\end{minipage}
\end{figure}

This is illustrated in Figure~\ref{Fig:NAPS}, where we show the mean signal integrated in 1-MHz band around the observation frequency for \Abell{1205} observations plotted versus the cycle number. Different observation dates are distinguished by different colors. On the top panel we have used as input the DAB-calibrated spectra obtained with standard NAPS tools. On the bottom panel we use as input the "raw spectra" and the calibration is done with the radio source \CalibSrc\, (from dedicated "drift scan" observations, in a similar way to that explained in Section~\ref{Sec:Calibration_3C161}). No data selection or RFI-cleaning algorithm is used in any case.

We see that the NAPS-calculated mean signal systematically increases from the beginning to the end of an observation, the maximum variation lies between approximatively 200~mJy and 1.2~Jy. This bias is not seen on the bottom panel, and it is clearly introduced by the DAB pulses. It is the main reason why we have done calibration using \CalibSrc\, observations and we have build a special pipeline to analyze the ACRT data. This is the same conclusion
as for the BAORadio pipeline (see Sec.~\ref{Sec:Calibration}).

Nevertheless, it is worth saying that the NRT pipeline and analysis tools and methods are optimized to search for known lines (frequency and width), and it is not adequate for the blind search we do in this study, as we have shown.



\newpage



\end{document}